\documentclass[12pt]{article}
\usepackage{amsmath,amssymb,amsthm,amsxtra,overpic,bbm,bm,epsfig,subfigure}
\usepackage{hyperref}
\usepackage{mathrsfs}
\usepackage{enumitem}
\usepackage{graphicx}
\usepackage{color}
\usepackage{comment}
\usepackage{epstopdf}
\usepackage{float}
\usepackage{cite}
\usepackage{slashed,stmaryrd}

\allowdisplaybreaks[1]

\def\thefootnote{\fnsymbol{footnote}}

\usepackage{multirow}
\newcommand{\Dlr}{\mbox{$\raisebox{2mm}{\boldmath ${}^\leftrightarrow$}\hspace{-4mm}D^{}_\mu$}}
\newcommand{\Dilr}{\mbox{$\raisebox{2mm}{\boldmath ${}^\leftrightarrow$}\hspace{-4mm}D^I_\mu$}}
\newcommand{\Dilrs}{\mbox{$\raisebox{2mm}{\boldmath ${}^\leftrightarrow$}\hspace{-4mm}\slashed{D}^I$}}
\newcommand{\Dlrs}{\mbox{$\raisebox{2mm}{\boldmath ${}^\leftrightarrow$}\hspace{-4mm}\slashed{D}$}}
\newcommand{\Dl}{\mbox{$\raisebox{2mm}{\boldmath ${}^\leftarrow$}\hspace{-4mm}D^{}_\mu$}}

\newcommand{\Yn}{Y^{}_\nu}
\newcommand{\DYn}{Y^\dagger_\nu}
\newcommand{\Yl}{Y^{}_l}
\newcommand{\DYl}{Y^\dagger_l}
\newcommand{\Yu}{Y^{}_{\rm u}}
\newcommand{\DYu}{Y^\dagger_{\rm u}}
\newcommand{\Yd}{Y^{}_{\rm d}}
\newcommand{\DYd}{Y^\dagger_{\rm d}}

\newcommand{\DDYni}[2][i]{\left(Y^\dagger_\nu Y^{}_\nu\right)^{}_{#1 #2}} % Y^\dagger_\nu Y^{}_\nu
\newcommand{\Yni}[2][\alpha]{\left(Y^{}_\nu\right)^{}_{#1 #2}} % Y^{}_\nu
\newcommand{\DYni}[2][\beta]{\left(Y^\dagger_\nu \right)^{}_{#2 #1}} % Y^\dagger_\nu
\newcommand{\TYni}[2][\beta]{\left(Y^{\rm T}_\nu \right)^{}_{#2 #1}}
\newcommand{\Yli}[2][\alpha]{\left(Y^{}_l\right)^{}_{#1 #2}}
\newcommand{\DYli}[2][\beta]{\left(Y^\dagger_l\right)^{}_{#1 #2}}
\newcommand{\Yui}[2][\alpha]{\left(Y^{}_{\rm u}\right)^{}_{#1 #2}}
\newcommand{\DYui}[2][\beta]{\left(Y^\dagger_{\rm u}\right)^{}_{#1 #2}}
\newcommand{\Ydi}[2][\alpha]{\left(Y^{}_{\rm d}\right)^{}_{#1 #2}}
\newcommand{\DYdi}[2][\beta]{\left(Y^\dagger_{\rm d}\right)^{}_{#1 #2}}

\newcommand{\Lelli}[1][\beta]{\ell^{}_{#1 \rm L}}
\newcommand{\BLelli}[1][\alpha]{\overline{\ell^{}_{#1 \rm L}}}
\newcommand{\REi}[1][\beta]{E^{}_{#1 \rm R}}
\newcommand{\BREi}[1][\alpha]{\overline{E^{}_{#1 \rm R}}}
\newcommand{\LQi}[1][\beta]{Q^{}_{#1 \rm L}}
\newcommand{\BLQi}[1][\alpha]{\overline{Q^{}_{#1 \rm L}}}
\newcommand{\RUi}[1][\beta]{U^{}_{#1 \rm R}}
\newcommand{\BRUi}[1][\alpha]{\overline{U^{}_{#1 \rm R}}}
\newcommand{\RDi}[1][\beta]{D^{}_{#1 \rm R}}
\newcommand{\BRDi}[1][\alpha]{\overline{D^{}_{#1 \rm R}}}

\newcommand{\rmI}{{\rm i}}
\newcommand{\lnmi}{ L^{}_i }
\newcommand{\lnmk}{ L^{}_k }
\newcommand{\lnik}{ L^{}_{ik} }
\newcommand{\lnkj}{ L^{}_{kj} }
\newcommand{\lnji}{ L^{}_{ji} }
\newcommand{\pM}[2][i]{M^{#2}_#1}
\newcommand{\Op}{\mathcal{O}}

\newcommand{\str}[1]{{\rm STr} \left( #1 \right)}
\newcommand{\strs}[1]{{\rm STr} \left[ #1 \right]}
\newcommand{\kit}[1]{K^{-1}_{#1}}
\newcommand{\xt}[1]{X^{}_{#1}}

\usepackage{ulem}

\begin{document}

\begin{center}
{\Large\bf Complete One-loop Matching of the Type-I Seesaw Model onto the Standard Model Effective Field Theory}
\end{center}

\vspace{0.2cm}

\begin{center}
{\bf Di Zhang~$^{a,~b}$}~\footnote{E-mail: zhangdi@ihep.ac.cn},
\quad
{\bf Shun Zhou~$^{a,~b}$}~\footnote{E-mail: zhoush@ihep.ac.cn (corresponding author)}
\\
\vspace{0.2cm}
{\small
$^a$Institute of High Energy Physics, Chinese Academy of Sciences, Beijing 100049, China\\
$^b$School of Physical Sciences, University of Chinese Academy of Sciences, Beijing 100049, China}
\end{center}

\vspace{1.5cm}

\begin{abstract}
In this paper, we accomplish the complete one-loop matching of the type-I seesaw model onto the Standard Model Effective Field Theory (SMEFT), by integrating out three heavy Majorana neutrinos with the functional approach. It turns out that only 31 dimension-six operators (barring flavor structures and Hermitian conjugates) in the Warsaw basis of the SMEFT can be obtained, and most of them appear at the one-loop level. The Wilson coefficients of these 31 dimension-six operators are computed up to $\mathcal{O}\left( M^{-2}\right)$ with $M$ being the mass scale of heavy Majorana neutrinos. As the effects of heavy Majorana neutrinos are encoded in the Wilson coefficients of these higher-dimensional operators, a complete one-loop matching is useful to explore the low-energy phenomenological consequences of the type-I seesaw model. In addition, the threshold corrections to the couplings in the Standard Model and to the coefficient of the dimension-five operator are also discussed.
\end{abstract}

%\begin{flushleft}
%\hspace{0.8cm} PACS number(s):
%\end{flushleft}

\def\thefootnote{\arabic{footnote}}
\setcounter{footnote}{0}
\newpage
%\tableofcontents

\section{Introduction}\label{sec:intro}

The Standard Model (SM) has been extremely successful in describing strong and electroweak interactions and passed almost all the tests of precision measurements~\cite{Zyla:2020zbs}. However, tiny neutrino masses and the particle candidates for dark matter cannot be accommodated in the SM, implying that the SM is actually incomplete and serves as only an effective field theory (EFT) at the low-energy scale. In the Standard Model Effective Field Theory (SMEFT)~\cite{Buchmuller:1985jz,Grzadkowski:2010es}, the effective operators of mass dimension higher than four composed of the SM fields are introduced and the SM gauge symmetry is preserved. Therefore, the low-energy observables computed in the SMEFT framework contain the effects of new physics but are independent of any specific ultraviolet (UV) model, offering a useful way to search for new physics beyond the SM.

Great progress in various aspects of the SMEFT has been made in recent years~\cite{Lehman:2014jma,Liao:2016hru,Li:2020gnx,Murphy:2020rsh,Li:2020xlh,Liao:2020jmn,Li:2020tsi,
Jenkins:2013zja,Jenkins:2013wua,Alonso:2013hga,Liao:2019tep,David:2020pzt,Henning:2015alf,Antusch:2001ck,Chala:2021juk} (see, e.g., Ref.~\cite{Brivio:2017vri}, for a comprehensive review) and several interesting extensions of the SMEFT (e.g., $\nu$SMEFT as the SMEFT extended with sterile neutrinos) have been developed~\cite{Aparici:2009fh, delAguila:2008ir, Bhattacharya:2015vja, Liao:2016qyd,Bischer:2019ttk,Li:2021tsq}. On the other hand, if a specific renormalizable UV model is known, one can match it onto the SMEFT by integrating out the heavy degrees of freedom to study its low-energy consequences. But, generally speaking, not all of the effective operators of mass dimension larger than four in the SMEFT can be induced. In Ref.~\cite{deBlas:2017xtg}, all the tree-level contributions to the Wilson coefficients of the dimension-six SMEFT operators in any UV completions with general scalar, spinor and vector field content and arbitrary interactions have been derived. However, the one-loop matching is in general more complicated and more technical than the tree-level matching. So far there have been only a few examples of complete or partial one-loop matching for the simple extensions of the SM, such as the SM extended with a charged scalar singlet~\cite{Bilenky:1993bt}, a real scalar singlet~\cite{Boggia:2016asg, Ellis:2017jns, Jiang:2018pbd, Haisch:2020ahr, Cohen:2020fcu,Dittmaier:2021fls}, a real scalar triplet~\cite{Henning:2016lyp, Ellis:2016enq, Fuentes-Martin:2016uol}, a vector-like quark singlet~\cite{delAguila:2016zcb}, a light sterile neutrino and heavy fermions and a scalar singlet~\cite{Chala:2020vqp}, two scalar leptoquarks~\cite{Gherardi:2020det}, and singlet right-handed neutrinos~\cite{Zhang:2021tsq} by either diagrammatic calculations or the functional approach~\cite{Henning:2014wua, Drozd:2015rsp, Ellis:2016enq, Henning:2016lyp, Fuentes-Martin:2016uol, Zhang:2016pja, DasBakshi:2018vni, Ellis:2017jns, Kramer:2019fwz, Cohen:2019btp, Cohen:2020fcu, Cohen:2020qvb, Fuentes-Martin:2020udw,Dittmaier:2021fls} (see, e.g., Refs.~\cite{Gaillard:1985uh,Chan:1986jq,Cheyette:1987qz}, for earlier works).

In this paper, we aim to carry out the complete one-loop matching for the effective operators up to dimension-six by integrating out heavy Majorana neutrinos in the canonical type-I seesaw model~\cite{Minkowski:1977sc, Yanagida:1979as, GellMann:1980vs, Glashow:1979nm, Mohapatra:1979ia} with the functional approach. The type-I seesaw model is the simplest and most natural one among various extensions of the SM to accommodate tiny neutrino masses~\cite{Xing:2019vks}. Thus it is important and necessary to explore its phenomenological consequences, especially at low-energy scales where precision measurements are performed. From the EFT point of view, the impact of the heavy Majorana neutrinos on low-energy observables is completely encoded in the Wilson coefficients of effective operators of dimension higher than four. This observation indicates that the complete one-loop matching is unavoidable to obtain the effective operators and the associated Wilson coefficients in the era of precision measurements. The tree-level matching for the type-I seesaw model has been performed before~\cite{Broncano:2002rw, Broncano:2003fq} and only one dimension-five operator and one dimension-six operator appear up to dimension-six (the operators of dimension-seven can be found in Ref.~\cite{Elgaard-Clausen:2017xkq}), where the former one is the unique Weinberg operator generating tiny Majorana masses of ordinary neutrinos~\cite{Weinberg:1979sa} and the latter one is a linear combination of two dimension-six operators in the Warsaw basis~\cite{Grzadkowski:2010es}. The dimension-six operator modifies the coupling of neutrinos with weak gauge bosons and then induces the unitarity violation of lepton flavor mixing~\cite{Broncano:2002rw, Broncano:2003fq,Broncano:2004tz,Antusch:2006vwa,Abada:2007ux,Antusch:2014woa,Elgaard-Clausen:2017xkq}. Nevertheless, the complete one-loop matching for the type-I seesaw model up to dimension-six is still lacking though partial results have been discussed in previous works~\cite{Zhang:2021tsq,Brivio:2018rzm,Coy:2018bxr}. This is the main motivation for a complete one-loop matching of the type-I seesaw model onto the SMEFT. Moreover, such calculations exemplify the complete one-loop matching for some UV models.

The functional approach based on the so-called covariant derivative expansion (CDE)~\cite{Gaillard:1985uh,Chan:1986jq,Cheyette:1987qz} will be utilized to perform the one-loop matching. Compared with diagrammatic calculations, the functional approach does not involve the evaluation of a large set of Green functions governed by corresponding Feynman diagrams both in the UV model and in the EFT, where a generic basis of operators in the EFT (i.e., the Green's basis~\cite{Jiang:2018pbd,Gherardi:2020det}) is required to be constructed first. With the functional approach, one can obtain directly the complete set of effective operators and the corresponding Wilson coefficients in the EFT without any prior knowledge of the EFT. Due to these advantages, the functional approach is more suitable for computer programming. Recently, two such {\sf Mathematica} packages, i.e., {\sf STrEAM}~\cite{Cohen:2020qvb} and {\sf SuperTracer}~\cite{Fuentes-Martin:2020udw}, have been made available publicly. Both of them follow the prescription proposed in Ref.~\cite{Cohen:2020fcu} and calculate the supertraces (which will be introduced in detail in Sec.~\ref{sec:framework}) by means of the CDE method. Since the package {\sf SuperTracer} allows for the substitution of interaction terms in a specific UV theory and partially simplifies the resulting operators, we adopt it to evaluate the relevant supertraces in the type-I seesaw model. The operators obtained by {\sf SuperTracer} are redundant, and one needs to apply algebraic, Fierz identities, integration by parts and the equations of motion (EOMs) of fields to convert these operators into those in the Warsaw basis. Such a procedure is cumbersome and gives threshold corrections to the coefficients of operators of dimension less than six, including those of the renormalizable operators in the SM~\cite{Wells:2017vla}. To make this procedure in order and the results traceable, we first reduce these operators into those in the Green's basis by using only algebraic, Fierz identities and integration by parts, then with the help of EOMs of fields convert them into the operators in the Warsaw basis. Besides the one-loop matching, we also carry out the tree-level matching for the type-I seesaw model and list the induced operators up to dimension-six, which have been obtained in Refs.~\cite{Broncano:2002rw, Broncano:2003fq}. Actually, the latter is not only for completeness but also necessary, since we also take into consideration the threshold corrections to the coefficients of the renormalizable and dimension-five operators, and the redefinitions of relevant fields give rise to additional one-loop contributions to the Wilson coefficients of two dimension-six operators in the Warsaw basis.

The remaining part of this paper is organized as follows. In Sec.~\ref{sec:framework}, we introduce the general formalism and explain how to make use of the functional approach to perform the tree-level and one-loop matchings for an UV model. In Sec.~\ref{sec:model}, we recall the type-I seesaw model and carry out the tree-level and one-loop matchings, where the operators appearing at the tree level together with the associated Wilson coefficients and the supertraces resulted from the one-loop matching are given. After evaluating the supertraces by using {\sf SuperTracer} and simplifying the results, we list all the operators and their Wilson coefficients, as well as the corrections to the couplings in the SM and the Wilson coefficient of the dimension-five operator in the Green's basis in Sec.~\ref{sec:Green}. In Sec~\ref{sec:Warsaw}, we present the corresponding results in the Warsaw basis which are converted from those in the Green's basis by applying the fields' EOMs, where the threshold corrections are also taken into account. Finally we summarize our main conclusions in Sec.~\ref{sec:summary}.

\section{Matching via the Functional Approach}\label{sec:framework}

In this section, we set up the framework of the functional approach to perform the tree-level and one-loop matchings between the low-energy EFT and an UV theory by integrating out the heavy degrees of freedom. This framework has been first presented in Ref.~\cite{Zhang:2016pja}. The main idea to match a given UV theory to the low-energy EFT is to equate the one-light-particle-irreducible (1LPI) effective action (i.e., $\Gamma^{}_{\rm L, UV}$) in the UV theory with the one-particle-irreducible (1PI) effective action (i.e., $\Gamma^{}_{\rm EFT}$) in the low-energy EFT at the matching scale, namely,
\begin{eqnarray}\label{eq:matching-condition}
\Gamma^{}_{\rm L, UV} \left[ \phi^{}_{\rm B} \right] = \Gamma^{}_{\rm EFT} \left[ \phi^{}_{\rm B} \right] \;,
\end{eqnarray}
where both effective actions are the functionals of the light background fields $\phi^{}_{\rm B}$ and can be calculated by using the background field method (see~\cite{Abbott:1981ke} for more details and earlier references). Here, for simplicity, we consider the case where the heavy fields $\Phi$ and the light fields $\phi$ in the UV theory are real scalar fields. For other types of fields, the relevant results can be easily generalized.

\subsection{Calculation of $\Gamma^{}_{\rm L, UV}$}

The generating functional of correlation functions in the UV theory is given by
\begin{eqnarray}\label{eq:gen-fun-corr}
Z^{}_{\rm UV} \left[ J^{}_\Phi, J^{}_\phi \right] = \int \mathcal{D} \Phi \mathcal{D} \phi \exp \left\{\rmI \int {\rm d}^d x \left( \mathcal{L}^{}_{\rm UV} \left[ \Phi, \phi \right] + J^{}_\Phi \Phi + J^{}_\phi \phi \right) \right\} \;,
\end{eqnarray}
in which $J^{}_\Phi$ and $J^{}_\phi$ are external sources for $\Phi$ and $\phi$, respectively, $d \equiv 4-2\varepsilon$ is the space-time dimension (i.e., we always work in the $d$-dimensional space-time), and the integrations over the heavy fields $\Phi$ and the light fields $\phi$ have been separated explicitly. One can split all heavy and light fields into classical background parts $\Phi^{}_{\rm B}$ and $\phi^{}_{\rm B}$, and quantum fluctuations $\Phi^\prime$ and $\phi^\prime$, namely
\begin{eqnarray}\label{eq:split}
\Phi = \Phi^{}_{\rm B} + \Phi^\prime \;, \qquad
\phi = \phi^{}_{\rm B} + \phi^\prime \;,
\end{eqnarray}
with the background fields satisfying the classical EOMs in the presence of external sources, i.e.,
\begin{eqnarray}\label{eq:cl-eom}
\frac{\delta \mathcal{L}^{}_{\rm UV}}{\delta \Phi} \left[ \Phi^{}_{\rm B}, \phi^{}_{\rm B} \right] + J^{}_\Phi &=& 0 \;,
\nonumber
\\
\frac{\delta \mathcal{L}^{}_{\rm UV}}{\delta \phi} \left[ \Phi^{}_{\rm B}, \phi^{}_{\rm B} \right] + J^{}_\phi &=& 0 \;.
\end{eqnarray}
With the help of Eqs.~(\ref{eq:split}) and (\ref{eq:cl-eom}), one can expand the Lagrangian of the UV theory $\mathcal{L}^{}_{\rm UV}$ together with the source terms around the classical background fields up to the second order of quantum fields, and thus obtain
\begin{eqnarray}\label{eq:expansion}
\mathcal{L}^{}_{\rm UV} \left[ \Phi, \phi \right] + J^{}_\Phi \Phi + J^{}_\phi \phi \simeq \mathcal{L}^{}_{\rm UV} \left[ \Phi^{}_{\rm B}, \phi^{}_{\rm B} \right] + J^{}_\Phi \Phi^{}_{\rm B} + J^{}_\phi \phi^{}_{\rm B} - \frac{1}{2} \left(\begin{matrix} \Phi^{\prime {\rm T}}  & \phi^{\prime {\rm T}} \end{matrix}\right) \mathcal{Q}^{}_{\rm UV} \left(\begin{matrix} \Phi^\prime \\ \phi^\prime \end{matrix} \right) \;,
\end{eqnarray}
where higher-order terms have been neglected and
\begin{eqnarray}\label{eq:ex-quadratic}
\mathcal{Q}^{}_{\rm UV}  \equiv \left(\begin{matrix} \displaystyle  -\frac{\delta^2 \mathcal{L}^{}_{\rm UV} }{\delta \Phi^2 } \left[ \Phi^{}_{\rm B}, \phi^{}_{\rm B} \right] &\hspace{0.3cm} \displaystyle -\frac{\delta^2 \mathcal{L}^{}_{\rm UV} }{\delta \Phi \delta \phi } \left[ \Phi^{}_{\rm B}, \phi^{}_{\rm B} \right]
\\
\displaystyle -\frac{\delta^2 \mathcal{L}^{}_{\rm UV} }{\delta \phi \delta \Phi } \left[ \Phi^{}_{\rm B}, \phi^{}_{\rm B} \right] &\hspace{0.3cm} \displaystyle -\frac{\delta^2 \mathcal{L}^{}_{\rm UV} }{ \delta \phi^2 } \left[ \Phi^{}_{\rm B}, \phi^{}_{\rm B} \right] \end{matrix} \right) \equiv \left(\begin{matrix} \Delta^{}_{\Phi} & X^{}_{\Phi \phi} \\ X^{}_{\phi \Phi} & \Delta^{}_{\phi} \end{matrix}\right) \;.
\end{eqnarray}
Notice that when there are several components of the heavy and light fields, $\Phi$ and $\phi$ (and the corresponding background and quantum fields) should be viewed as column vectors. Then, by making use of Eqs.~(\ref{eq:split}) and ~(\ref{eq:expansion}), we can recast the generating functional in Eq.~(\ref{eq:gen-fun-corr}) into the following approximate form
\begin{eqnarray}\label{eq:approx-gf}
Z^{}_{\rm UV} \left[ J^{}_\Phi, J^{}_\phi \right] &\simeq& \exp \left\{ \rmI \int {\rm d}^d x \left( \mathcal{L}^{}_{\rm UV} \left[ \Phi^{}_{\rm B}, \phi^{}_{\rm B} \right] + J^{}_\Phi \Phi^{}_{\rm B} + J^{}_\phi \phi^{}_{\rm B} \right) \right\}
\nonumber
\\
&& \times \int \mathcal{D} \Phi^\prime \mathcal{D} \phi^\prime \exp \left\{ -\frac{\rmI}{2} \int {\rm d}^d x \left(\begin{matrix} \Phi^{\prime {\rm T}}  & \phi^{\prime {\rm T}} \end{matrix}\right) \mathcal{Q}^{}_{\rm UV} \left(\begin{matrix} \Phi^\prime \\ \phi^\prime \end{matrix} \right)  \right\}
\nonumber
\\
&\propto&  \exp \left\{ \rmI \int {\rm d}^d x \left( \mathcal{L}^{}_{\rm UV} \left[ \Phi^{}_{\rm B}, \phi^{}_{\rm B} \right] + J^{}_\Phi \Phi^{}_{\rm B} + J^{}_\phi \phi^{}_{\rm B} \right) \right\} \times \left( \det \mathcal{Q}^{}_{\rm UV} \right)^{-c^{}_s} \;,
\end{eqnarray}
with $c^{}_{\rm s}$ accounting for the spin statistics and the number of degrees of freedom of the fields integrated over. For instance, $c^{}_{\rm s} = 1/2$ for real bosonic fields in the case under consideration, $c^{}_{\rm s} = 1$ for complex bosonic fields, and $c^{}_{\rm s} = -1$ (or $-1/2$) for Dirac (or Majorana) fermionic fields.

The 1LPI effective action $\Gamma^{}_{\rm L, UV}\left[ \phi^{}_{\rm B} \right]$ is defined as the Legendre transformation of the generating functional of connected correlation functions with $J^{}_\Phi = 0$, i.e.,
\begin{eqnarray}\label{eq:1LPI-action}
\Gamma^{}_{\rm L, UV} \left[ \phi^{}_{\rm B} \right] &\equiv& -\rmI \ln Z^{}_{\rm UV}\left[J^{}_\Phi=0, J^{}_\phi \right] - \int {\rm d}^d
x J^{}_\phi \phi^{}_{\rm B}
\nonumber
\\
&\simeq& \int {\rm d}^d x \mathcal{L}^{}_{\rm UV} \left[ \Phi^{}_{\rm c} \left[ \phi^{}_{\rm B} \right], \phi^{}_{\rm B} \right] + \frac{\rmI}{2} \ln \det \mathcal{Q}^{}_{\rm UV} \left[ \Phi^{}_{\rm c} \left[ \phi^{}_{\rm B} \right], \phi^{}_{\rm B} \right] \;,
\end{eqnarray}
where Eq.~(\ref{eq:approx-gf}) with $c^{}_{\rm s} = 1/2$ is taken into account in the last step, and the classical heavy field $\Phi^{}_{\rm c} \left[ \phi^{}_{\rm B} \right] \equiv \Phi^{}_{\rm B} \left[ J^{}_\Phi = 0, J^{}_\phi \right]$ satisfying
\begin{eqnarray}\label{eq:cl-eom-Phi}
\left.\frac{\delta \mathcal{L}^{}_{\rm UV} \left[ \Phi, \phi \right] }{ \delta \Phi } \right|^{}_{\Phi = \Phi^{}_{\rm c} \left[ \phi^{}_{\rm B} \right] ,\, \phi = \phi^{}_{\rm B}} = 0 \;.
\end{eqnarray}
The first and second terms in the last line of Eq.~(\ref{eq:1LPI-action}) stand for the tree- and one-loop-level contributions to $\Gamma^{}_{\rm L, UV}\left[ \phi^{}_{\rm B} \right]$, respectively. But at this point, it should be noticed that $\Gamma^{}_{\rm L,UV} \left[ \phi^{}_{\rm B} \right]$ is in general non-local due to the non-locality of $\Phi^{}_{\rm c} \left[ \phi^{}_{\rm B} \right]$ determined by Eq.~(\ref{eq:cl-eom-Phi}). To obtain the local functionals, one can expand $\Phi^{}_{\rm c} \left[ \phi^{}_{\rm B} \right]$ to a given order in $1/M$ and denote the local one by $\widehat{\Phi}^{}_{\rm c} \left[ \phi^{}_{\rm B} \right]$, where $M$ represents the mass scale of heavy fields that is considered to be extremely high. Substituting $\widehat{\Phi}^{}_{\rm c} \left[ \phi^{}_{\rm B} \right]$ into Eq.~(\ref{eq:1LPI-action}), one obtains the local $\Gamma^{}_{\rm L,UV} \left[ \phi^{}_{\rm B} \right]$ at a given order in $1/M$, i.e.,
\begin{eqnarray}\label{eq:1LPI-split}
\Gamma^{\rm tree}_{\rm L, UV} \left[ \phi^{}_{\rm B} \right] &=& \int {\rm d}^d x \mathcal{L}^{}_{\rm UV} \left[ \widehat{\Phi}^{}_{\rm c} \left[ \phi^{}_{\rm B} \right], \phi^{}_{\rm B} \right] \;,
\nonumber
\\
\Gamma^{\rm 1-loop}_{\rm L, UV} \left[ \phi^{}_{\rm B} \right] &=& \frac{\rmI}{2} \ln \det \mathcal{Q}^{}_{\rm UV} \left[ \widehat{\Phi}^{}_{\rm c} \left[ \phi^{}_{\rm B} \right], \phi^{}_{\rm B} \right] \;.
\end{eqnarray}
It is worthwhile to mention that the separation of the tree- and one-loop-level contributions and the loop counting in the UV theory are unambiguous.

\subsection{Calculation of $\Gamma^{}_{\rm EFT}$}
As the matching between the UV theory and the low-energy EFT is performed order by order (i.e., that in Eq.~(\ref{eq:matching-condition})), one can formally split the effective Lagrangian of the EFT also into the tree- and loop-level parts,
\begin{eqnarray}\label{eq:eft-Lagrangian}
\mathcal{L}^{}_{\rm EFT} \left[ \phi \right]  \simeq \mathcal{L}^{\rm tree}_{\rm EFT} \left[ \phi \right] + \mathcal{L}^{\rm 1-loop}_{\rm EFT} \left[ \phi \right] \;,
\end{eqnarray}
in which the higher-loop contributions have been neglected. In Eq.~(\ref{eq:eft-Lagrangian}), $\mathcal{L}^{\rm tree}_{\rm EFT} \left[ \phi \right] $ and $\mathcal{L}^{\rm 1-loop}_{\rm EFT} \left[ \phi \right]$ are composed of tree- and one-loop-level effective operators and the associated Wilson coefficients, respectively. Similar to the computation of $\Gamma^{}_{\rm L, UV}$, one can easily derive the generating functional of correlation functions up to one-loop level in the low-energy EFT
\begin{eqnarray}\label{eq:gen-fun-corr-eft}
Z^{}_{\rm EFT} \left[ J^{}_\phi \right] &=& \int \mathcal{D} \phi \exp \left\{ \rmI \int {\rm d}^d x \left( \mathcal{L}^{}_{\rm EFT} \left[ \phi \right] + J^{}_\phi \phi \right) \right\}
\nonumber
\\
&\propto& \exp \left\{ \rmI \int {\rm d}^d x \left( \mathcal{L}^{\rm tree}_{\rm EFT} \left[ \phi^{}_{\rm B} \right] + \mathcal{L}^{\rm 1-loop}_{\rm EFT} \left[ \phi^{}_{\rm B} \right] + J^{}_\phi \phi^{}_{\rm B} \right) \right\} \times \left( \det \mathcal{Q}^{}_{\rm EFT} \right)^{-1/2} \;,
\end{eqnarray}
with
\begin{eqnarray}\label{eq:ex-quadratic-eft}
\mathcal{Q}^{}_{\rm EFT} \equiv -\frac{\delta^2 \mathcal{L}^{\rm tree}_{\rm EFT} }{\delta \phi^2} \left[ \phi^{}_{\rm B} \right] \;,
\end{eqnarray}
where only the tree-level Lagrangian is applied to calculate $\mathcal{Q}^{}_{\rm EFT}$ since the contributions from the one-loop part result in a higher-order functional determinant. With the help of Eq.~(\ref{eq:gen-fun-corr-eft}), we can derive the 1PI effective action $\Gamma^{}_{\rm EFT}$ up to the one-loop order, viz.
\begin{eqnarray}\label{eq:1LPI-action-eft}
\Gamma^{}_{\rm EFT} \left[ \phi^{}_{\rm B} \right] &=& - \rmI \ln Z^{}_{\rm EFT} \left[ J^{}_\phi \right] - \int {\rm d}^d x J^{}_\phi \phi^{}_{\rm B}
\nonumber
\\
&\simeq& \int {\rm d}^d x \left( \mathcal{L}^{\rm tree}_{\rm EFT} \left[ \phi^{}_{\rm B} \right] + \mathcal{L}^{\rm 1-loop}_{\rm EFT} \left[ \phi^{}_{\rm B} \right] \right) + \frac{\rmI}{2} \ln \det \mathcal{Q}^{}_{\rm EFT} \;,
\end{eqnarray}
whose tree-level and one-loop parts can be clearly separated,
\begin{eqnarray}\label{eq:1LPI-split-eft}
\Gamma^{\rm tree}_{\rm EFT} \left[ \phi^{}_{\rm B} \right] &=& \int {\rm d}^d x  \mathcal{L}^{\rm tree}_{\rm EFT} \left[ \phi^{}_{\rm B} \right] \;,
\nonumber
\\
\Gamma^{\rm 1-loop}_{\rm EFT} \left[ \phi^{}_{\rm B} \right] &=& \int {\rm d}^d x \mathcal{L}^{\rm 1-loop}_{\rm EFT} \left[ \phi^{}_{\rm B} \right] + \frac{\rmI}{2} \ln \det \mathcal{Q}^{}_{\rm EFT} \;.
\end{eqnarray}

It is now evident that the one-loop effective action of the EFT consists of two different parts. One is the one-loop effective operators contained in $\mathcal{L}^{\rm 1-loop}_{\rm EFT}$, and the other one is contained in $\ln \det \mathcal{Q}^{}_{\rm EFT}$, taking account of the contributions from the tree-level effective Lagrangian $\mathcal{L}^{\rm tree}_{\rm EFT}$ via one-loop corrections.

\subsection{Matching}\label{sec:matching}

With the help of Eqs.~(\ref{eq:matching-condition}), (\ref{eq:1LPI-split}) and (\ref{eq:1LPI-split-eft}), one can accomplish the matching order by order and then arrive at
\begin{eqnarray}\label{eq:tree-matching}
\mathcal{L}^{\rm tree}_{\rm EFT} \left[ \phi^{}_{\rm B} \right] = \mathcal{L}^{}_{\rm UV} \left[ \widehat{\Phi}^{}_{\rm c} \left[ \phi^{}_{\rm B} \right], \phi^{}_{\rm B} \right] \;,
\end{eqnarray}
and
\begin{eqnarray}\label{eq:loop-matching}
\int {\rm d}^d x \mathcal{L}^{\rm 1-loop}_{\rm EFT} \left[ \phi^{}_{\rm B} \right] + \frac{\rmI}{2} \ln \det \mathcal{Q}^{}_{\rm EFT} \left[ \phi^{}_{\rm B} \right] = \frac{\rmI}{2} \ln \det \mathcal{Q}^{}_{\rm UV} \left[ \widehat{\Phi}^{}_{\rm c} \left[ \phi^{}_{\rm B} \right], \phi^{}_{\rm B} \right] \;,
\end{eqnarray}
where $\widehat{\Phi}^{}_{\rm c} \left[ \phi^{}_{\rm B} \right]$ satisfies the classical EOM given in Eq.~(\ref{eq:cl-eom-Phi}) and has been localized. As shown in Eq.~(\ref{eq:tree-matching}), the tree-level matching can be performed by simply substituting the localized solutions of the classical EOMs for the heavy fields into the Lagrangian of the UV theory. To find out the one-loop Lagrangian of the EFT, we have to first deal with the second term on the left-hand side of Eq.~(\ref{eq:loop-matching}), namely,
\begin{eqnarray}\label{eq:Q-eft1}
\mathcal{Q}^{}_{\rm EFT} \left[ \phi \right] &=& - \frac{\delta^2 \mathcal{L}^{\rm tree}_{\rm EFT}\left[ \phi\right] }{\delta \phi^2}  = - \frac{\delta}{\delta \phi} \left( \frac{\delta \mathcal{L}^{}_{\rm UV} \left[ \widehat{\Phi}^{}_{\rm c} \left[ \phi \right], \phi \right] }{\delta \phi} \right)
\nonumber
\\
&=& -\frac{\delta}{\delta \phi} \left( \frac{\delta \mathcal{L}^{}_{\rm UV} }{\delta \phi} [ \widehat{\Phi}^{}_{\rm c} \left[ \phi \right], \phi ] + \frac{\delta \widehat{\Phi}^{}_{\rm c} \left[ \phi \right] }{\delta \phi} \frac{ \delta \mathcal{L}^{}_{\rm UV} }{ \delta \Phi} \left[ \widehat{\Phi}^{}_{\rm c} \left[ \phi \right], \phi \right] \right)
\nonumber
\\
&=& - \frac{\delta^2 \mathcal{L}^{}_{\rm UV}}{ \delta \phi^2} \left[ \widehat{\Phi}^{}_{\rm c} \left[ \phi \right], \phi \right] - \frac{\delta \widehat{\Phi}^{}_{\rm c} \left[ \phi \right]}{\delta \phi} \frac{\delta^2 \mathcal{L}^{}_{\rm UV} }{\delta \Phi \delta \phi} \left[ \widehat{\Phi}^{}_{\rm c} \left[ \phi \right], \phi \right]
\nonumber
\\
&=& \Delta^{}_{\phi} \left[ \widehat{\Phi}^{}_{\rm c} \left[ \phi \right], \phi \right] + \frac{\delta \widehat{\Phi}^{}_{\rm c} \left[ \phi \right]}{\delta \phi} X^{}_{\Phi \phi} \left[ \widehat{\Phi}^{}_{\rm c} \left[ \phi \right], \phi \right] \;,
\end{eqnarray}
where Eqs.~(\ref{eq:ex-quadratic}) and (\ref{eq:cl-eom-Phi}) have been used. Then, considering the identity
\begin{eqnarray}\label{eq:Q-eft2}
0 &=& \frac{\delta}{ \delta \phi } \left( \frac{\delta \mathcal{L}^{}_{\rm UV }}{\delta \Phi} \left[ \widehat{\Phi}^{}_{\rm c} \left[ \phi \right], \phi \right] \right) = \frac{\delta^2 \mathcal{L}^{}_{\rm UV}}{ \delta \phi \delta \Phi} \left[ \widehat{\Phi}^{}_{\rm c} \left[ \phi \right],\phi\right] + \frac{\delta \widehat{\Phi}^{}_{\rm c}\left[ \phi \right] }{\delta \phi} \frac{\delta^2 \mathcal{L}^{}_{\rm UV}}{\delta \Phi^2} \left[ \widehat{\Phi}^{}_{\rm c} \left[ \phi \right], \phi \right]
\nonumber
\\
&=& X^{}_{\phi\Phi} \left[ \widehat{\Phi}^{}_{\rm c} \left[ \phi \right], \phi \right] + \frac{\delta \widehat{\Phi}^{}_{\rm c} \left[ \phi \right] }{\delta \phi} \Delta^{}_{\Phi} \left[ \widehat{\Phi}^{}_{\rm c} \left[ \phi \right], \phi \right] \;,
\end{eqnarray}
we have
\begin{eqnarray}\label{eq:Q-eft3}
\frac{\delta \widehat{\Phi}^{}_{\rm c} \left[ \phi \right]}{\delta \phi}  = - \left( X^{}_{\phi \Phi} \widehat{\Delta}^{-1}_{\Phi} \right) \left[ \widehat{\Phi}^{}_{\rm c} \left[ \phi \right], \phi \right] \;,
\end{eqnarray}
in which $\widehat{\Delta}^{-1}_\Phi$ is the local expansion of $\Delta^{-1}_\Phi$. Inserting Eq.~(\ref{eq:Q-eft3}) into Eq.~(\ref{eq:Q-eft1}), one gets
\begin{eqnarray}\label{eq:Q-eft4}
\frac{\rmI}{2} \ln\det \mathcal{Q}^{}_{\rm EFT} \left[ \phi \right] &=& \frac{\rmI}{2} \ln\det \left( \Delta^{}_\phi - X^{}_{\phi \Phi} \widehat{\Delta}^{-1}_\Phi X^{}_{\Phi \phi} \right) \left[ \widehat{\Phi}^{}_{\rm c} \left[ \phi \right], \phi \right]
\nonumber
\\
&=& \frac{\rmI}{2} \ln\det \Delta^{}_\phi \left[ \widehat{\Phi}^{}_{\rm c} \left[ \phi \right], \phi \right] + \frac{\rmI}{2} \ln\det \left( 1 - \Delta^{-1}_\phi X^{}_{\phi\Phi} \widehat{\Delta}^{-1}_{\Phi} X^{}_{\Phi\phi} \right) \left[ \widehat{\Phi}^{}_{\rm c} \left[ \phi \right], \phi \right]
\nonumber
\\
&=& \frac{\rmI}{2} \ln\det \Delta^{}_\phi \left[ \widehat{\Phi}^{}_{\rm c} \left[ \phi \right], \phi \right] + \frac{\rmI}{2} \ln\det \left( 1 - \widehat{\Delta}^{-1}_{\Phi} X^{}_{\Phi\phi} \Delta^{-1}_\phi X^{}_{\phi\Phi} \right) \left[ \widehat{\Phi}^{}_{\rm c} \left[ \phi \right], \phi \right]
\nonumber
\\
&=& \frac{\rmI}{2} \ln\det \Delta^{}_\phi \left[ \widehat{\Phi}^{}_{\rm c} \left[ \phi \right], \phi \right] - \frac{\rmI}{2} \ln\det \widehat{\Delta}^{}_\Phi \left[ \widehat{\Phi}^{}_{\rm c} \left[ \phi \right], \phi \right]
\nonumber
\\
&&+ \frac{\rmI}{2} \ln\det \left( \widehat{\Delta}^{}_\Phi -  X^{}_{\Phi\phi} \Delta^{-1}_\phi X^{}_{\phi\Phi} \right) \left[ \widehat{\Phi}^{}_{\rm c} \left[ \phi \right], \phi \right] \;.
\end{eqnarray}
Note that $\mathcal{Q}^{}_{\rm UV}$ on the right-hand side of Eq.~(\ref{eq:loop-matching}) can be block-diagonalized by the transformation with unit Jacobian determinant~\cite{Fuentes-Martin:2016uol}, i.e.,
\begin{eqnarray}\label{eq:Q-uv1}
\left( \begin{matrix} 1 &\hspace{0.3cm} 0 \\ -\Delta^{-1}_\phi X^{}_{\phi \Phi} &\hspace{0.3cm} 1  \end{matrix} \right)^\dagger \mathcal{Q}^{}_{\rm UV} \left( \begin{matrix} 1 &\hspace{0.3cm} 0 \\ -\Delta^{-1}_\phi X^{}_{\phi \Phi} &\hspace{0.3cm} 1  \end{matrix} \right) = \left( \begin{matrix} \Delta^{}_\Phi - X^{}_{\Phi\phi} \Delta^{-1}_\phi X^{}_{\phi \Phi} &\hspace{0.3cm} 0 \\ 0 &\hspace{0.3cm} \Delta^{}_\phi \end{matrix} \right) \;,
\end{eqnarray}
with which one can further obtain
\begin{eqnarray}\label{eq:Q-uv2}
\frac{\rmI}{2} \ln \det \mathcal{Q}^{}_{\rm UV} \left[ \widehat{\Phi}^{}_{\rm c} \left[ \phi \right], \phi \right] &=& \frac{\rmI}{2} \ln \det \left( \Delta^{}_\Phi - X^{}_{\Phi\phi} \Delta^{-1}_\phi X^{}_{\phi\Phi} \right) \left[ \widehat{\Phi}^{}_{\rm c} \left[ \phi \right], \phi \right] \nonumber \\
&~& + \frac{\rmI}{2} \ln \det \Delta^{}_\phi \left[ \widehat{\Phi}^{}_{\rm c} \left[ \phi \right], \phi \right] \;.
\end{eqnarray}
Substituting Eqs.~(\ref{eq:Q-eft4}) and (\ref{eq:Q-uv2}) into Eq.~(\ref{eq:loop-matching}), we have
\begin{eqnarray}\label{eq:loop-matching1}
\int {\rm d}^d x \mathcal{L}^{\rm 1-loop}_{\rm EFT} \left[ \phi^{}_{\rm B} \right] &=& \frac{\rmI}{2} \ln \det \left( \Delta^{}_\Phi - X^{}_{\Phi\phi} \Delta^{-1}_\phi X^{}_{\phi\Phi} \right) \left[ \widehat{\Phi}^{}_{\rm c} \left[ \phi^{}_{\rm B} \right], \phi^{}_{\rm B} \right] + \frac{\rmI}{2} \ln \det \widehat{\Delta}^{}_\Phi \left[ \widehat{\Phi}^{}_{\rm c} \left[ \phi^{}_{\rm B} \right], \phi^{}_{\rm B} \right]
\nonumber
\\
&& - \frac{\rmI}{2} \ln \det \left( \widehat{\Delta}^{}_\Phi - X^{}_{\Phi\phi} \Delta^{-1}_\phi X^{}_{\phi\Phi} \right) \left[ \widehat{\Phi}^{}_{\rm c} \left[ \phi^{}_{\rm B} \right], \phi^{}_{\rm B} \right] \;.
\end{eqnarray}
As expected, the contributions from pure light loops contained in $\ln \det \Delta^{}_\phi$ are cancelled out.

Usually, the calculations of the functional determinants in Eq.~(\ref{eq:loop-matching1}) involve the loop integrals. To calculate these loop integrals, one can implement the method of expansion by regions~\cite{Beneke:1997zp, Smirnov:2002pj, Jantzen:2011nz}. More explicitly, the loop integrals can be split into hard- (i.e., $k\sim M \gg p$ with $k$ and $p$ standing for the loop and external momenta, respectively) and soft-momentum (i.e., $k \sim p \ll M$) regions in the dimensional regularization with the modified minimal subtraction ($\rm \overline{MS}$) scheme. In each region, the integrand of the loop integral is expanded as a Taylor series with respect to the parameters that are considered small, and then the integrand should be integrated over the whole $d$-dimensional space of the loop momentum (i.e., $k$). In this way, the functional determinants can be separated into the so-called hard and soft parts as the loop integrals, leading to
\begin{eqnarray}\label{eq:expansion-by-region}
&&\ln \det \left( \Delta^{}_\Phi - X^{}_{\Phi\phi} \Delta^{-1}_\phi X^{}_{\phi\Phi} \right) = \ln \det \left( \Delta^{}_\Phi - X^{}_{\Phi\phi} \Delta^{-1}_\phi X^{}_{\phi\Phi} \right) |^{}_{\rm hard} + \ln \det \left( \Delta^{}_\Phi - X^{}_{\Phi\phi} \Delta^{-1}_\phi X^{}_{\phi\Phi} \right) |^{}_{\rm soft} \;,
\nonumber
\\
&&\ln\det \left( \widehat{\Delta}^{}_\Phi - X^{}_{\Phi\phi} \Delta^{-1}_\phi X^{}_{\phi\Phi} \right) = \ln \det \left( \Delta^{}_\Phi - X^{}_{\Phi\phi} \Delta^{-1}_\phi X^{}_{\phi\Phi} \right) |^{}_{\rm soft} \;,
\nonumber
\\
&&\ln\det \widehat{\Delta}^{}_\Phi = \ln\det \Delta^{}_\Phi |^{}_{\rm soft} = 0 \;.
\end{eqnarray}
In the last line of Eq.~(\ref{eq:expansion-by-region}), one should notice the fact that $\ln \det \Delta^{}_\Phi$ involves only heavy field propagators (i.e., pure heavy loops) and its soft part gives rise to scaleless integrals, which are vanishing in the end. Plugging Eq.~(\ref{eq:expansion-by-region}) into Eq.~(\ref{eq:loop-matching1}), we can obtain
\begin{eqnarray}\label{eq:loop-matching2}
\int {\rm d}^d x \mathcal{L}^{\rm 1-loop}_{\rm EFT} \left[ \phi^{}_{\rm B} \right] =\left. \frac{\rmI}{2} \ln \det \left( \Delta^{}_\Phi - X^{}_{\Phi\phi} \Delta^{-1}_\phi X^{}_{\phi\Phi} \right) \left[ \widehat{\Phi}^{}_{\rm c} \left[ \phi^{}_{\rm B} \right], \phi^{}_{\rm B} \right] \right|^{}_{\rm hard} \;.
\end{eqnarray}
Starting with Eq.~(\ref{eq:loop-matching2}), one can directly carry out the one-loop matching by calculating the hard part of $\ln \det \left( \Delta^{}_\Phi - X^{}_{\Phi\phi} \Delta^{-1}_\phi X^{}_{\phi\Phi} \right)$, as described in Ref.~\cite{Fuentes-Martin:2016uol}.

However, there is an alternative way that is more suitable to be implemented in a computer program~\cite{Cohen:2020qvb,Fuentes-Martin:2020udw}. Considering that $\ln\det \Delta^{}_\phi$ contains pure light loops and then its hard part is scaleless (i.e., vanishing after integration), then by virtue of Eqs.~(\ref{eq:1LPI-split}), (\ref{eq:Q-uv2}) and (\ref{eq:loop-matching2}), we have
\begin{eqnarray}\label{eq:loop-matching3}
\int {\rm d}^d x \mathcal{L}^{\rm 1-loop}_{\rm EFT} \left[ \phi^{}_{\rm B} \right] = \left. \Gamma^{\rm 1-loop}_{\rm L, UV} \left[ \phi^{}_{\rm B} \right] \right|^{}_{\rm hard} = \left. \frac{\rmI}{2} \ln \det \mathcal{Q}^{}_{\rm UV} \left[ \widehat{\Phi}^{}_{\rm c} \left[ \phi^{}_{\rm B} \right], \phi^{}_{\rm B} \right] \right|^{}_{\rm hard} \;.
\end{eqnarray}
Generally speaking, $\mathcal{Q}^{}_{\rm UV}$ can be decomposed into an inverse-propagator part $\bm{K}$ and an interaction part $\bm{X}$, such that~\cite{Cohen:2020fcu}
\begin{eqnarray}\label{eq:loop-matching4}
\int {\rm d}^d x \mathcal{L}^{\rm 1-loop}_{\rm EFT} \left[ \phi \right] &=& \left. \frac{\rmI}{2} \ln {\rm Sdet} \left( - \bm{K} + \bm{X} \right) \right|^{}_{\rm hard} = \left. \frac{\rmI}{2} {\rm STr} \ln \left( - \bm{K } + \bm{X} \right) \right|^{}_{\rm hard}
\nonumber
\\
&=& \left. \frac{\rmI}{2} {\rm STr} \ln \left( - \bm{K} \right) \right|^{}_{\rm hard} - \left. \frac{\rmI}{2} \sum^{\infty}_{n=1} \frac{1}{n} {\rm STr} \left[ \left( \bm{K}^{-1} \bm{X} \right)^n \right] \right|^{}_{\rm hard} \;,
\end{eqnarray}
where the superdeterminant ``Sdet'' denotes the generalization of the regular determinant providing fermionic blocks with an inverse power while leaving bosonic blocks as usual, and similarly, the supertrace ``STr'' is the generalization of the trace over both the internal degrees of freedom and the functional space giving fermionic blocks a minus sign. Moreover, in the last step in Eq.~(\ref{eq:loop-matching4}), the fact that $\bm{K}^{-1} \bm{X} \sim M^{-1}$~\cite{Fuentes-Martin:2020udw} has been taken into account to expand the logarithmic term, i.e., $\ln \left( \bm{1} - \bm{K}^{-1} \bm{X} \right)$. As one can see in Eq.~(\ref{eq:loop-matching4}), there are two types of terms constituting the one-loop EFT Lagrangian: the {\it log-type} and {\it power-type} supertraces corresponding to the first and second terms in the last line of Eq.~(\ref{eq:loop-matching4}), respectively. Some comments are in order.
\begin{itemize}
\item The {\it log-type} supertrace is universal and depends on the representations of heavy particles in the UV theory, since it comes from the kinetic and mass terms and only those of the heavy particles contribute to the log-type term. Therefore, the log-type term generally leads to the pure gauge-field operators. Nevertheless, if the heavy particles are singlets in the UV theory, the log-type term is trivial and will not induce any operators at all.

\item The {\it power-type} terms depend on the interactions of both heavy and light particles contained in the interaction part $\bm{X}$. Thanks to the relation $\bm{K}^{-1} \bm{X} \sim M^{-1}$ in the hard-momentum region, the power-type terms can be truncated according to the desired order of $1/M$ in the EFT Lagrangian, so the number of these power-type terms will be finite.
\end{itemize}

In order to gain the inverse-propagator part $\bm{K}$ and the interaction part $\bm{X}$ in Eq.~(\ref{eq:loop-matching4}), one has to calculate the second functional derivative of the Lagrangian with respect to all fields and their conjugates as well for complex fields in the UV theory. Hence it is quite useful to arrange the fields together their conjugates into field multiplets:
\begin{eqnarray}\label{eq:multiplet}
\varphi^{}_S = \left( \begin{matrix} S \\ S^\ast \end{matrix} \right) \;,\quad \varphi^{}_F = \left( \begin{matrix} F \\ F^c \end{matrix} \right) \;,\quad \varphi^{}_V = \left( \begin{matrix} V^{}_\mu \\ V^\ast_\mu \end{matrix} \right) \;,
\end{eqnarray}
for complex scalars, Dirac fermions and complex vectors, respectively, where $F^c \equiv {\sf C} \overline{F}^{\rm T}$ with ${\sf C}\equiv {\rm i}\gamma^2\gamma^0$ being the charge-conjugation matrix. For later convenience, one can define $\overline{\varphi}^{}_S = \varphi^\dagger_S$, $\overline{\varphi}^{}_V = \varphi^\dagger_V$ and $\overline{\varphi}^{}_F = \varphi^\dagger_F \gamma^0$ for the conjugated field multiplets. In addition, if $S$ (or $V$) is a real scalar (or vector) and $F$ is a Majorana fermion, then we have $\varphi^{}_{S (V)} = S(V)$, $\overline{\varphi}^{}_{S(V)} = S^{\rm T} (V^{\rm T})$ and $\varphi^{}_{F} = F$, $\overline{\varphi}^{}_F = \overline{F}$. Consequently, $\bm{K}$ and $\bm{X}$ can be extracted via
\begin{eqnarray}\label{eq:derivative}
\left. \delta^2 \mathcal{L}^{}_{\rm UV} \right|^{}_{\Phi = \widehat{\Phi}^{}_{\rm c} \left[ \phi \right] } = 2 \left. \mathcal{L}^{}_{\rm UV} \left[ \varphi + \delta \varphi \right] \right|^{}_{\Phi = \widehat{\Phi}^{}_{\rm c} \left[ \phi \right] } \supset \delta \overline{\varphi}^{}_i \left( \bm{K}^{}_i \delta^{}_{ij} - \bm{X}^{}_{ij} \right) \delta \varphi^{}_j \;,
\end{eqnarray}
where only the relevant terms $\mathcal{O} \left( \delta \varphi^2 \right) $ are retained and $i,j= 1,...,n^{}_S, n^{}_S + 1,..., n^{}_S+ n^{}_V, n^{}_S+ n^{}_V +1,..., n^{}_S+ n^{}_V + n^{}_F$ with $n^{}_S$, $n^{}_V$ and $n^{}_F$ being the numbers of scalars, vectors and fermions. The inverse-propagator part $\bm{K}$ is a block-diagonal matrix and its sub-blocks are
\begin{eqnarray}\label{eq:K-part}
\bm{K}^{}_i = \left\{ \begin{array}{ll} P^2- m^2_i &\hspace{0.5cm} \left({\rm scalar}\right) \\ \slashed{P} - m^{}_i &\hspace{0.5cm} \left( {\rm fermion} \right) \\  -g^{\mu\nu} (P^2 - m^2_i) + (1 - \xi^{-1}) P^\mu P^\nu &\hspace{0.5cm} \left( {\rm vector} \right) \end{array} \right. \;,
\end{eqnarray}
where $P^{}_\mu \equiv \rmI D^{}_\mu $ with $D^{}_\mu$ being the covariant derivative. Here, as in the background field approach, one can choose the Feynman-'t Hooft gauge for the quantum gauge fields (i.e., $\xi =1$ in the vector propagators) but still the general $R^{}_\xi$ gauge for the classical gauge fields. As for the interaction part $\bm{X}$, they can be generally cast into the form~\cite{Fuentes-Martin:2020udw}
\begin{eqnarray}\label{eq:X-part}
\bm{X} \left( P^{}_\mu, \phi \right) = \sum^{\infty}_{n=0} X^{\mu^{}_1 \cdots \mu^{}_n}_n \left( \phi \right) P^{}_{\mu^{}_1} \cdots P^{}_{\mu^{}_n} \;,
\end{eqnarray}
where all the ``open" covariant derivatives $P^{}_\mu$ are arranged to be rightmost and $X^{\mu^{}_1 \cdots \mu^{}_n}_n(\phi)$ are actually functionals of light fields, while the ``closed" covariant derivatives are usually written as the commutators, e.g., $\left[ P^{}_\mu, \phi \right] = \rmI \left( D^{}_\mu \phi \right)$.

After the inverse-propagator part $\bm{K}$ and the interaction part $\bm{X}$ in the UV theory are obtained with the help of Eq.~(\ref{eq:derivative}), the functional supertraces given in Eq.~(\ref{eq:loop-matching4}) can be evaluated by means of the CDE method~\cite{Gaillard:1985uh,Chan:1986jq,Cheyette:1987qz}. The technical details of the CDE method can be found in Refs.~\cite{Henning:2014wua,Cohen:2019btp,Fuentes-Martin:2020udw}. The aforementioned {\sf Mathematica} packages, i.e., {\sf SuperTracer}~\cite{Fuentes-Martin:2020udw} and {\sf STrEAM}~\cite{Cohen:2020fcu}, have implemented the CDE method to evaluate all functional supertraces appearing in Eq.~(\ref{eq:loop-matching4}). In the present work, we utilize the package {\sf SuperTracer} to calculate the functional supertraces, as it allows for the substitution of the $\bm{X}$ interactions from a specific theory and can partially simplify the resulting operators.

\section{The Type-I Seesaw Model}\label{sec:model}

As emphasized in Sec.~\ref{sec:intro}, the canonical type-I seesaw model~\cite{Minkowski:1977sc, Yanagida:1979as, GellMann:1980vs, Glashow:1979nm, Mohapatra:1979ia} is a natural extension of the SM to generate tiny neutrino masses. In this model, three right-handed neutrino singlets $N^{}_{\rm R}$ are introduced and a huge Majorana mass term for them is allowed. Another salient feature of this model is to offer an elegant explanation for the cosmological matter-antimatter asymmetry via the CP-violating and out-of-equilibrium decays of heavy Majorana neutrinos\cite{Fukugita:1986hr}. More explicitly, the Lagrangian of the type-I seesaw model is given by
\begin{eqnarray}\label{eq:Lagrangian}
\mathcal{L}^{}_{\rm UV} = \mathcal{L}^{}_{\rm SM} + \overline{N^{}_{\rm R}} \rmI \slashed{\partial} N^{}_{\rm R} - \left( \frac{1}{2} \overline{N^c_{\rm R}} M N^{}_{\rm R} + \overline{\ell^{}_{\rm L}} Y^{}_\nu \widetilde{H} N^{}_{\rm R} + {\rm h.c.} \right) \;,
\end{eqnarray}
with the SM Lagrangian
\begin{eqnarray}\label{eq:Lagrangian-SM}
\mathcal{L}^{}_{\rm SM} &=& -\frac{1}{4} G^A_{\mu\nu} G^{A\mu\nu} -\frac{1}{4} W^I_{\mu\nu} W^{I\mu\nu} -\frac{1}{4} B^{}_{\mu\nu} B^{\mu\nu}
\nonumber
\\
&&+ \sum^{}_f \overline{f} \rmI \slashed{D} f - \left( \overline{Q^{}_{\rm}} Y^{}_{\rm u} \widetilde{H} U^{}_{\rm R} + \overline{Q^{}_{\rm L}} Y^{}_{\rm d} H D^{}_{\rm R} + \overline{\ell^{}_{\rm L}} Y^{}_l H E^{}_{\rm R} + {\rm h.c.} \right)
\nonumber
\\
&& + \left( D^{}_\mu H \right)^\dagger \left( D^\mu H \right) - m^2 H^\dagger H - \lambda \left( H^\dagger H \right)^2 \;,
\end{eqnarray}
where $f=Q^{}_{\rm L}, U^{}_{\rm R}, D^{}_{\rm R}, \ell^{}_{\rm L}, E^{}_{\rm R}$ refer to the SM fermionic doublets and singlets, the covariant derivative $D^{}_\mu \equiv \partial^{}_\mu - \rmI g^{}_1 Y B^{}_\mu - \rmI g^{}_2 T^I W^I_\mu - \rmI g^{}_s \lambda^A G^A_\mu$ has been defined, and the other notations should be self-evident. In the basis where the Majorana mass matrix of right-handed neutrinos is diagonal, i.e., $M = {\rm Diag}\{M^{}_1, M^{}_2, M^{}_3\}$, one can obtain the EOMs of the heavy Majorana neutrinos from Eq.~(\ref{eq:Lagrangian}), namely,
\begin{eqnarray} \label{eq:EOM-N}
\left( \rmI \slashed{\partial} - M \right) N - \left( Y^\dagger_\nu \widetilde{H}^\dagger \ell^{}_{\rm L} + Y^{\rm T}_\nu \widetilde{H}^{\rm T} \ell^c_{\rm L} \right) = 0 \;,
\end{eqnarray}
where $N \equiv N^c_{\rm R} + N^{}_{\rm R}$ denotes heavy Majorana neutrinos in the mass basis. The solution for above EOMs leads us to
\begin{eqnarray}\label{eq:classical-N}
N \simeq - \left( M^{-1} + M^{-2} \rmI \slashed{\partial}\right) \left( Y^\dagger_\nu \widetilde{H}^\dagger \ell^{}_{\rm L} + Y^{\rm T}_\nu \widetilde{H}^{\rm T} \ell^c_{\rm L} \right) \;,
\end{eqnarray}
where only the terms up to $\mathcal{O}\left(M^{-2}\right)$ are kept since we are concerned about the effective operators up to dimension six in the low-energy EFT, namely, the seesaw EFT (SEFT).

\subsection{Tree-level matching}
Guided by the matching condition in Eq.~(\ref{eq:tree-matching}), one can easily obtain the tree-level part of the Lagrangian in the SEFT by substituting Eq.~(\ref{eq:classical-N}) into Eq.~(\ref{eq:Lagrangian}), i.e.,
\begin{eqnarray}
\mathcal{L}^{\rm tree}_{\rm SEFT} = \mathcal{L}^{}_{\rm SM} + \left[ \frac{1}{2} C^{(5)}_{\alpha\beta} \mathcal{O}^{(5)}_{\alpha\beta} + {\rm h.c.} \right] + C^{(6)}_{\alpha\beta} \mathcal{O}^{(6)}_{\alpha\beta} \;,
\end{eqnarray}
with
\begin{eqnarray}\label{eq:tree-operator}
\mathcal{O}^{(5)}_{\alpha\beta} = \BLelli \widetilde{H} \widetilde{H}^{\rm T} \ell^c_{\beta \rm L} \;,\quad \mathcal{O}^{(6)}_{\alpha\beta} = \left( \BLelli \widetilde{H} \right) \rmI \slashed{\partial} \left( \widetilde{H}^\dagger \Lelli \right) \;,
\end{eqnarray}
where the corresponding Wilson coefficients are $C^{(5)}_{\alpha\beta} = \left( Y^{}_\nu M^{-1} Y^{\rm T}_\nu \right)^{}_{\alpha\beta} $ and $C^{(6)}_{\alpha\beta} = \left( Y^{}_\nu M^{-2} Y^\dagger_\nu \right)^{}_{\alpha\beta} $. Therefore, the tree-level matching only induces two effective operators up to dimension six, coinciding with the results in the previous works~\cite{Broncano:2002rw,Broncano:2003fq}. One is the unique dimension-five operator $\mathcal{O}^{(5)}$, or the Weinberg operator, generating the Majorana masses of the ordinary neutrinos after the spontaneous gauge symmetry breaking~\cite{Weinberg:1979sa}. The other one is a dimension-six operator which modifies the couplings of neutrinos to gauge bosons and then causes the unitarity violation of the lepton flavor mixing matrix. The dimension-six operator $\mathcal{O}^{(6)}$ can be rewritten as a combination of two dimension-six operators in the Warsaw basis of the SMEFT~\cite{Grzadkowski:2010es}:
\begin{eqnarray} \label{eq:6-tree}
\mathcal{O}^{(6)}_{\alpha\beta} = \frac{1}{4} \left[ \left( \BLelli \gamma^\mu \Lelli \right) \left( H^\dagger \rmI \Dlr H \right) - \left( \BLelli \gamma^\mu \tau^I \Lelli \right) \left( H^\dagger \rmI \Dilr H \right) \right] \;,
\end{eqnarray}
where $ \Dlr \equiv D^{}_\mu - \Dl$ and $\Dilr \equiv \tau^I D^{}_\mu - \Dl \tau^I$ with $\tau^I$ (for $I=1,2,3$) being the Pauli matrices and $\Dl$ acting on the left.

\subsection{One-loop matching}
To proceed with the one-loop matching, one needs to calculate the matrices $\bm{K}$ and $\bm{X}$ in the type-I seesaw model. In this case, the field multiplet $\varphi$ can be identified as
\begin{eqnarray}
\varphi^{}_i \in \{ \varphi^{}_N,\; \varphi^{}_\ell,\; \varphi^{}_E,\; \varphi^{}_Q,\; \varphi^{}_U,\; \varphi^{}_D,\; \varphi^{}_H,\; \varphi^{}_W,\; \varphi^{}_B \} \;,
\end{eqnarray}
where
\begin{eqnarray}
\varphi^{}_N = N \;,\quad \varphi^{}_F = \left(\begin{matrix} F \\ F^c \end{matrix} \right) \;,\quad \varphi^{}_H = \left( \begin{matrix} H \\  H^* \end{matrix} \right) \;,\quad \varphi^{}_V = V \;,
\end{eqnarray}
with $F =\ell, E, Q, U, D$ and $V=W, B$. Notice that here the ${\rm SU}(3)$ gauge fields in the SM and all ghost fields have been omitted, since they do not take part in the one-loop matching by integrating the heavy Majorana neutrinos. The sub-blocks of the inverse-propagator matrix $\bm{K}$ take the form with entries given in Eq.~(\ref{eq:K-part}). It is worth pointing out that the inverse propagators of the heavy Majorana neutrinos, i.e., $\left( \rmI \slashed{\partial} - M^{}_i \right)$, do not involve any gauge bosons. This means that the corresponding log-type supertrace in Eq.~(\ref{eq:loop-matching4}) gives only a constant and does not lead to any operators in the SEFT. As for the entries of the interaction matrix $\bm{X}$, they can be extracted by making use of Eq.~(\ref{eq:derivative}) and written in the form given in Eq.~(\ref{eq:X-part}):
\begin{align}\label{eq:X-N}
X^{}_{HN} &= \left( \begin{matrix} \overline{\ell^{}_{\rm L}} Y^{}_\nu \epsilon \\ \overline{\ell^c_{\rm L}} Y^\ast_\nu \epsilon  \end{matrix} \right) \;,\quad & X^{}_{NH} &= \left(\begin{matrix} \epsilon^{\rm T} Y^\dagger_\nu \ell^{}_{\rm L} &\hspace{0.2cm} \epsilon^{\rm T} Y^{\rm T}_\nu \ell^c_{\rm L} \end{matrix} \right) \;,
\nonumber
\\
X^{}_{\ell N} &= \left( \begin{matrix} Y^{}_\nu \widetilde{H}  P^{}_{\rm R} \\ Y^\ast_\nu \widetilde{H}^\ast  P^{}_{\rm L} \end{matrix} \right) \;,\quad & X^{}_{N\ell} &= \left( \begin{matrix} \widetilde{H}^\dagger Y^\dagger_\nu P^{}_{\rm L} &\hspace{0.2cm} \widetilde{H}^{\rm T} Y^{\rm T}_\nu P^{}_{\rm R}  \end{matrix} \right) \;,
\nonumber
\\
X^{}_{\ell H} & = \left(\begin{matrix} Y^{}_l E^{}_{\rm R} &\hspace{0.2cm} \epsilon Y^{}_\nu P^{}_{\rm R} N \\ \epsilon Y^\ast_\nu P^{}_{\rm L} N &\hspace{0.2cm} Y^\ast_l E^c_{\rm R} \end{matrix} \right) \;,\quad &\hspace{-2cm} X^{}_{H\ell} &= \left(\begin{matrix} \overline{E^{}_{\rm R}} Y^\dagger_l &\hspace{0.2cm} \epsilon^{\rm T} \overline{N} Y^{\rm T}_\nu P^{}_{\rm R} \\ \epsilon^{\rm T} \overline{N} Y^\dagger_\nu P^{}_{\rm L} &\hspace{0.2cm} \overline{E^c_{\rm R}} Y^{\rm T}_l \end{matrix}\right) \;,
\end{align}
and
\begin{eqnarray}\label{eq:X-SM}
X^{}_{HH} &=& \left(\begin{matrix} m^2 + 2\lambda \left( \left|H\right|^2 +  H H^\dagger\right) & 2\lambda H H^{\rm T} \\ 2\lambda H^\ast H^\dagger & m^2 + 2\lambda \left( \left|H\right|^2 + H^\ast H^{\rm T} \right)  \end{matrix}\right)
\nonumber
\\
X^\nu_{HB} &=& \rmI g^{}_1 \left(\begin{matrix} - \left(D^{}_\nu H\right) \\ \left( D^{}_\mu H \right)^\ast  \end{matrix}\right) + \frac{g^{}_1}{2} \left(\begin{matrix} -H \\ H^\ast \end{matrix}\right) \rmI D^{}_\nu \;,
\nonumber
\\
X^{\mu}_{BH} &=& \frac{\rmI g^{}_1}{2} \left(\begin{matrix} \left(D^{}_\mu H \right)^\dagger &\hspace{0.2cm} -\left(D^{}_\mu H \right)^{\rm T} \end{matrix} \right) + \frac{g^{}_1}{2} \left(\begin{matrix} -H^\dagger &\hspace{0.2cm} H^{\rm T} \end{matrix}\right) \rmI D^{}_\mu \;,
\nonumber
\\
X^{\nu J}_{HW} &=& \rmI g^{}_2 \left(\begin{matrix} - \tau^J \left(D^{}_\nu H\right) \\ \tau^{J\ast} \left(D^{}_\nu H\right)^\ast  \end{matrix} \right) + \frac{g^{}_2}{2} \left(\begin{matrix} - \tau^J H \\ \tau^{J\ast} H^\ast \end{matrix} \right) \rmI D^{}_\nu \;,
\nonumber
\\
X^{\mu I}_{WH} &=&  \frac{\rmI g^{}_2}{2} \left(\begin{matrix} \left( D^{}_\mu H \right)^\dagger \tau^I &\hspace{0.2cm} - \left(D^{}_\mu H \right)^{\rm T} \tau^{I\ast} \end{matrix} \right) + \frac{g^{}_2}{2} \left(\begin{matrix} -H^\dagger \tau^I &\hspace{0.2cm} H^{\rm T} \tau^{I \ast} \end{matrix}\right) \rmI D^{}_\mu \;.
\end{eqnarray}
Note that $\epsilon$ is the two-dimensional antisymmetric tensor with $\epsilon^{}_{12} = -\epsilon^{}_{21} = 1$, where the indices of this tensor are referring to the weak isospin space. The $X$ terms in Eq.~(\ref{eq:X-N}) (partly) denote the new contributions from the interactions of heavy Majorana neutrinos, and the heavy Majorana neutrino field $N$ in $X$ terms should be replaced by the terms on the right-hand side of Eq.~(\ref{eq:classical-N}). The corresponding $X$ terms for the SM interactions can be found in the Appendix B of Ref.~\cite{Cohen:2020fcu}, whereas the $X$ terms for the interactions between the Higgs boson and gauge bosons, as well as those among the Higgs bosons, should take the forms given in Eq.~(\ref{eq:X-SM}) due to different conventions used in the package {\sf SuperTracer}~\cite{Fuentes-Martin:2020udw} and in Ref.~\cite{Cohen:2020fcu}. In addition, different conventions of the quartic Higgs coupling in the literature should be noted.

With the inverse-propagator matrix $\bm{K}$ in Eq.~(\ref{eq:K-part}), as well as the interaction matrix $\bm{X}$ given in Eqs.~(\ref{eq:X-N}), (\ref{eq:X-SM}) and the Appendix B of Ref.~\cite{Cohen:2020fcu}, up to dimension six, we can use Eq.~(\ref{eq:loop-matching4}) to derive the effective Lagrangian
\begin{eqnarray}\label{eq:supertraces}
\int {\rm d}^d x \mathcal{L}^{\rm 1-loop}_{\rm EFT} &=&  - \frac{\rmI}{2} \left\{\vphantom{\frac{1}{1}}\hspace{-0.6mm} \str{\kit{N} \xt{NH} \kit{H} \xt{H N}} + \str{\kit{N} \xt{N\psi} \kit{\psi} \xt{\psi N}} \right.
\nonumber
\\
&& + \str{\kit{N} \xt{NH} \kit{H} \xt{HH} \kit{H} \xt{HN}} + \str{\kit{N} \xt{N\psi} \kit{\psi} \xt{\psi\psi} \kit{\psi} \xt{\psi N}}
\nonumber
\\
&& + \str{\kit{N} \xt{NH} \kit{H} \xt{HV} \kit{V} \xt{V H} \kit{H} \xt{HN}}
\nonumber
\\
&& + \str{ \kit{N} \xt{NH} \kit{H} \xt{H\psi} \kit{\psi} \xt{\psi H} \kit{H} \xt{H N} }
\nonumber
\\
&& + \str{ \kit{N} \xt{N\psi} \kit{\psi} \xt{\psi V} \kit{V} \xt{V\psi} \kit{\psi} \xt{\psi N} }
\nonumber
\\
&& +\str{ \kit{N} \xt{N\psi} \kit{\psi} \xt{\psi H} \kit{H} \xt{H \psi} \kit{\psi} \xt{\psi N} }
\nonumber
\\
&& + \str{\kit{N} \xt{N\psi} \kit{\psi} \xt{\psi\psi} \kit{\psi} \xt{\psi\psi} \kit{\psi} \xt{\psi N}}
\nonumber
\\
&& + \str{\kit{N} \xt{N\psi} \kit{\psi} \xt{\psi\psi} \kit{\psi} \xt{\psi\psi} \kit{\psi} \xt{\psi\psi} \kit{\psi} \xt{\psi N}}
\nonumber
\\
&& + \str{\kit{N} \xt{N\psi} \kit{\psi} \xt{\psi\psi} \kit{\psi} \xt{\psi\psi} \kit{\psi} \xt{\psi\psi} \kit{\psi} \xt{\psi\psi} \kit{\psi} \xt{\psi N}}
\nonumber
\\
&& + \left[ \str{\kit{N} \xt{NH} \kit{H} \xt{H\psi} \kit{\psi} \xt{\psi N}}  \right.
\nonumber
\\
&& + \str{\kit{N} \xt{NH} \kit{H} \xt{HV} \kit{V} \xt{V\psi} \kit{\psi} \xt{\psi N}}
\nonumber
\\
&& + \str{ \kit{N} \xt{NH} \kit{H} \xt{HH} \kit{H} \xt{H\psi} \kit{\psi} \xt{\psi N} }
\nonumber
\\
&&+ \str{ \kit{N} \xt{NH} \kit{H} \xt{H\psi} \kit{\psi} \xt{\psi\psi} \kit{\psi} \xt{\psi N} }
\nonumber
\\
&& + \str{\kit{N} \xt{NH} \kit{H} \xt{HV} \kit{V} \xt{VH} \kit{H} \xt{H\psi} \kit{\psi} \xt{\psi N}}
\nonumber
\\
&& + \str{\kit{N} \xt{NH} \kit{H} \xt{HV} \kit{V} \xt{V\psi} \kit{\psi} \xt{\psi\psi} \kit{\psi} \xt{\psi N}}
\nonumber
\\
&& + \str{\kit{N} \xt{NH} \kit{H} \xt{H\psi} \kit{\psi} \xt{\psi\psi} \kit{\psi} \xt{\psi\psi} \kit{\psi} \xt{\psi N}}
\nonumber
\\
&& + \str{\kit{N} \xt{N\psi} \kit{\psi} \xt{\psi V} \kit{V} \xt{VH} \kit{H} \xt{H\psi} \kit{\psi} \xt{\psi N}}
\nonumber
\\
&& + \str{\kit{N} \xt{N\psi} \kit{\psi} \xt{\psi V} \kit{V} \xt{V\psi} \kit{\psi} \xt{\psi\psi} \kit{\psi} \xt{\psi N}}
\nonumber
\\
&& + \left.\left.\left. \str{\kit{N} \xt{N\psi} \kit{\psi} \xt{\psi H} \kit{H} \xt{H\psi} \kit{\psi} \xt{\psi\psi} \kit{\psi} \xt{\psi N}} + {\rm h.c.} \right] \vphantom{\frac{1}{1}}\hspace{-0.6mm}\right\} \right|^{}_{\rm hard}
\nonumber
\\
&& - \frac{\rmI}{2} \left\{ \frac{1}{2} \strs{ \left(\kit{N} \xt{NH} \kit{H} \xt{HN} \right)^2  } + \frac{1}{2} \strs{ \left( \kit{N} \xt{N\psi} \kit{\psi} \xt{\psi N} \right)^2 } \right.
\nonumber
\\
&& + \frac{1}{2} \strs{ \left( \kit{N} \xt{N\psi} \kit{\psi} \xt{\psi\psi} \kit{\psi} \xt{\psi N} \right)^2 } + \frac{1}{3} \strs{ \left( \kit{N} \xt{N\psi} \kit{\psi} \xt{\psi N} \right)^3 }
\nonumber
\\
&& + \str{\kit{N} \xt{N\psi} \kit{\psi} \xt{\psi N} \kit{N} \xt{NH} \kit{H} \xt{H N}}
\nonumber
\\
&& + \str{\kit{N} \xt{N\psi} \kit{\psi} \xt{\psi \psi} \kit{\psi} \xt{\psi N} \kit{N} \xt{N H} \kit{H} \xt{H N}}
\nonumber
\\
&& + \str{\kit{N} \xt{N\psi} \kit{\psi} \xt{\psi N} \kit{N} \xt{N\psi} \kit{\psi} \xt{\psi\psi} \kit{\psi} \xt{\psi N}}
\nonumber
\\
&& + \str{\kit{N} \xt{N\psi} \kit{\psi} \xt{\psi N} \kit{N} \xt{N\psi} \kit{\psi} \xt{\psi\psi} \kit{\psi} \xt{\psi\psi} \kit{\psi} \xt{\psi N}}
\nonumber
\\
&& + \left. \left. \left[ \str{\kit{N} \xt{N\psi} \kit{\psi} \xt{\psi N} \kit{N} \xt{NH} \kit{H} \xt{H\psi} \kit{\psi} \xt{\psi N}} + {\rm h.c.} \right] \vphantom{\frac{1}{1}}\hspace{-0.6mm} \right\} \right|^{}_{\rm hard} \;,
\end{eqnarray}
where $\psi = \left( \ell, E, Q, U, D \right)$ and $V= \left( W, B \right)$ collectively denote the SM fermions and weak gauge bosons. It is worthwhile to stress that the terms in the second braces in Eq.~(\ref{eq:supertraces}) involve at least two heavy-neutrino propagators, so $1/2$ and $1/3$ in front of some supertraces are actually the symmetry factors counting the power of repeated blocks in a given supertrace. Therefore, in order to find out the effective operators at the one-loop level in the SEFT, one has to evaluate the long list of supertraces in Eq.~(\ref{eq:supertraces}), which is obviously tedious but can be achieved by using the package {\sf SuperTracer} in a straightforward way.

\section{Green's Basis}\label{sec:Green}

After the supertraces in Eq.~(\ref{eq:supertraces}) are calculated by using the package {\sf SuperTracer}, one needs to further convert the resultant operators into the independent operators in the Warsaw basis of the SMEFT. But, before doing so, we first convert these operators into those in a redundant basis, the so-called Green's basis~\cite{Jiang:2018pbd,Gherardi:2020det}, by taking advantage of algebraic, Fierz identities and integration by parts. Then, applying the EOMs of fields, we can obtain the independent operators in the Warsaw basis from those in the Green's basis.

\subsection{Threshold corrections}

Generally, the one-loop matching can result in threshold corrections to the renormalizable terms already existing in the SM, i.e.,
\begin{eqnarray}\label{eq:correction-GB}
\delta \mathcal{L} &=& \delta Z^{}_{G} G^A_{\mu\nu} G^{A\mu\nu} + \delta Z^{}_W W^I_{\mu\nu} W^{I\mu\nu} + \delta Z^{}_B B^{}_{\mu\nu} B^{\mu\nu}
\nonumber
\\
&&+ \sum^{}_f  \overline{f} \delta Z^{}_f \rmI \slashed{D} f + \left( \overline{Q^{}_{\rm}} \delta Y^{}_{\rm u} \widetilde{H} U^{}_{\rm R} + \overline{Q^{}_{\rm L}} \delta Y^{}_{\rm d} H D^{}_{\rm R} + \overline{\ell^{}_{\rm L}} \delta Y^{}_l H E^{}_{\rm R} + {\rm h.c.} \right)
\nonumber
\\
&& + \delta Z^{}_H \left( D^{}_\mu H \right)^\dagger \left( D^\mu H \right) + \delta m^2 H^\dagger H + \delta \lambda \left( H^\dagger H \right)^2 \;.
\end{eqnarray}
where $f=Q^{}_{\rm L}, U^{}_{\rm R}, D^{}_{\rm R}, \ell^{}_{\rm L}, E^{}_{\rm R}$. For a given UV model, not all the above terms are induced, which is of course dependent on the interactions of the heavy fields. In the type-I seesaw model, the one-loop matching leads to
\begin{eqnarray}\label{eq:renormalizable}
\left( 4\pi \right)^2 \delta Z^{\rm G}_H &=&  \frac{1}{2} \DDYni{i} \left( 1 + 2 \lnmi \right) \;,
\nonumber
\\
\left( 4\pi \right)^2 \left( \delta Z^{\rm G}_\ell \right)^{}_{\alpha\beta} &=& \frac{1}{4} \Yni{i} \DYni{i} \left( 1 + 2\frac{m^2}{M^2_i} \right) \left( 3 + 2 \lnmi \right) \;,
\nonumber
\\
\left( 4\pi \right)^2 \left( \delta m^2 \right)^{\rm G} &=& -2 \DDYni{i} \pM{2} \left( 1 + \lnmi \right) \;,
\nonumber
\\
\left( 4\pi \right)^2 \delta \lambda^{\rm G} &=& - \DDYni{k} \DDYni[k]{i} \frac{ \left( 1 + \lnmi \right) \pM{2} - \left( 1 + \lnmk \right) \pM[k]{2}}{\pM{2} - \pM[k]{2}}
\nonumber
\\
&& + \DDYni{k} \DDYni{k} \frac{\pM{} \pM[k]{} \lnik}{\pM{2}- \pM[k]{2}} \;,
\nonumber
\\
\left( 4\pi \right)^2 \left( \delta Y^{\rm G}_\ell \right)^{}_{\alpha\beta} &=& - \Yni{i} \left( \DYn \Yl \right)^{}_{i\beta} \left( 1 + \frac{m^2}{M^2_i} \right) \left( 1 + \lnmi \right)
\end{eqnarray}
with
\begin{eqnarray}\label{eq:log-def}
L^{}_i \equiv \ln \left( \frac{\mu^2}{\pM{2}} \right) + \frac{1}{\varepsilon} - \gamma^{}_{\rm E} + \ln \left( 4\pi \right) \;,\quad L^{}_{ij} \equiv \ln \left( \frac{\pM{2}}{\pM[j]{2}} \right) \; ,
\end{eqnarray}
up to $\mathcal{O}\left( M^{-2} \right)$ in the Green's basis, where the other terms in Eq.~(\ref{eq:correction-GB}) do not appear, the superscript ``G" signifies the results in the Green's basis, and $\gamma^{}_{\rm E}$ in Eq. (\ref{eq:log-def}) is the Euler constant. It is worth pointing out that the divergences in $L^{}_i$ come from the hard part of loop integrals and usually consist of both the UV and infrared (IR) divergences in the dimensional regularization. The UV divergences can be absorbed by the renormalization constants in the UV model but with heavy fields replaced with their classical EOMs, such as Eq.~(\ref{eq:classical-N}) in the type-I seesaw model, while the IR divergences can be regarded as part of the counterterms of the EFT to cancel corresponding UV divergences of the EFT~\cite{Dittmaier:2021fls}. Here, one can simply remove the $1/\varepsilon -\gamma^{}_{\rm E} + \ln \left( 4\pi \right)$ terms in $L^{}_i$ to get the renormalized couplings or Wilson coefficients in the $\overline{\rm MS}$ scheme~\cite{Bilenky:1993bt,Dittmaier:2021fls}.

In addition, there also exist one-loop corrections to the Wilson coefficient of the dimension-five operator $1/2 \left(\delta C^{(5)} \right)^{\rm G}_{\alpha\beta} \Op^{(5)}_{\alpha\beta}$ in Eq.~(\ref{eq:tree-operator}), namely,
\begin{eqnarray}
\left( 4\pi \right)^2 \left( \delta C^{(5)} \right)^{\rm G}_{\alpha\beta} = \left[ 2\lambda \left( 1 + \lnmi \right) + \frac{g^2_1+g^2_2}{4} \left( 1+ 3 \lnmi\right) \right] \Yni{i} \pM{-1} \TYni{i} \;.
\end{eqnarray}
This result will be important for us to find the threshold correction to the Wilson coefficient of the Weinberg operator in the Warsaw basis.

\subsection{Dimension-six operators}

%%%%%%%%%%%%%%%%%%%%%%%%%%%%% table 1 %%%%%%%%%%%%%%%%%%%%%%%%%%%%%%%%%%%
\begin{table}[!t]
\centering
\renewcommand\arraystretch{1.5}
\resizebox{\textwidth}{!}{
\begin{tabular}{c|c|c|c|c|c}
\hline\hline
\multicolumn{2}{c}{$X^2H^2$} & \multicolumn{2}{|c|}{$H^2D^4$} & \multicolumn{2}{c}{$H^6$}
\\
\hline
$\Op^{}_{HB}$ & $B^{}_{\mu\nu} B^{\mu\nu} H^\dagger H$ & $\Op^{}_{DH}$ & $\left( D^{}_\mu D^\mu H \right)^\dagger \left(D^{}_\nu D^\nu H \right)$ & $\Op^{}_H$ & $ \left( H^\dagger H\right)^3 $
\\
\cline{3-6}
$\Op^{}_{HW}$ & $W^I_{\mu\nu} W^{I\mu\nu} H^\dagger H$ & \multicolumn{2}{c|}{$H^4D^2$} & \multicolumn{2}{c}{$\psi^2 D^3$}
\\
\cline{3-6}
$\Op^{}_{HWB}$ & $W^I_{\mu\nu} B^{\mu\nu} \left( H^\dagger \tau^I H \right)$ & $ \Op^{}_{H \Box}$ & $\left( H^\dagger H \right) \Box \left( H^\dagger H \right) $ & $\Op^{\alpha\beta}_{\ell D}$ & $ \displaystyle \frac{\rmI}{2} \BLelli \left( D^2 \slashed{D} + \slashed{D} D^2 \right) \Lelli$
\\
\cline{1-2}\cline{5-6}
\multicolumn{2}{c|}{$H^2 X D^2$} & $\Op^{}_{HD}$ & $\left(H^\dagger D^{}_\mu H \right)^\ast \left(H^\dagger D^\mu H \right)$ & \multicolumn{2}{c}{$\psi^2 X H$}
\\
\cline{1-2}\cline{5-6}
$\Op^{}_{BDH}$ & $ D^{}_\nu B^{\mu\nu} \left(H^\dagger \rmI \Dlr H \right)$ & $\Op^\prime_{HD}$ & $\left( H^\dagger H \right) \left( D^{}_\mu H \right)^\dagger \left(D^\mu H \right)$ & $\Op^{\alpha\beta}_{eB} $ & $\left( \BLelli \sigma^{\mu\nu} \REi \right) H B^{}_{\mu\nu}$
\\
$\Op^{}_{WDH}$ & $\left(D^{}_\nu W^{\mu\nu} \right)^I  \left( H^\dagger \rmI \Dilr H \right)$ & $\Op^{\prime\prime}_{HD}$ & $ \left( H^\dagger H\right) D^\mu \left( H^\dagger \rmI \Dlr H \right)$ & $\Op^{\alpha\beta}_{eW}$ & $ \left( \BLelli \sigma^{\mu\nu} \REi \right) \tau^I H W^I_{\mu\nu}$
\\
\hline
\multicolumn{2}{c}{$\psi^2 H D^2$ } & \multicolumn{2}{|c|}{$\psi^2 D H^2$} & \multicolumn{2}{c}{$\psi^2 H^3$ }
\\
\hline
$\Op^{\alpha\beta}_{eHD1}$ & $\BLelli \REi D^2 H$ & $\Op^{(1)\alpha\beta}_{H\ell}$ & $\left( \BLelli \gamma^\mu \Lelli \right) \left( H^\dagger \rmI \Dlr H \right)$ & $\Op^{\alpha\beta}_{eH}$ & $ \left( \BLelli H \REi \right) \left(H^\dagger H \right) $
\\
\cline{5-6}
$\Op^{\alpha\beta}_{eHD2}$ & $ \BLelli \rmI \sigma^{\mu\nu} D^{}_\mu \REi D^{}_\nu H$ & $\Op^{\prime(1) \alpha\beta}_{H\ell}$ & $\left( \BLelli \rmI \Dlrs \Lelli \right) \left( H^\dagger H \right)$ & \multicolumn{2}{c}{Four-lepton}
\\
\cline{5-6}
$\Op^{\alpha\beta}_{eHD3}$ & $ \BLelli D^2 \REi H$ & $\Op^{\prime\prime(1)\alpha\beta}_{H\ell}$ & $ \left( \BLelli \gamma^\mu \Lelli \right) D^{}_\mu \left( H^\dagger H \right) $ & $\Op^{\alpha\beta\gamma\lambda}_{\ell\ell}$ & $\left( \BLelli \gamma^\mu \Lelli \right) \left( \BLelli[\gamma] \gamma^{}_\mu \Lelli[\lambda] \right)$
\\
$\Op^{\alpha\beta}_{eHD4}$ & $\BLelli D^\mu \REi D^{}_\mu H$ & $\Op^{(3)\alpha\beta}_{H\ell}$ & $\left( \BLelli \gamma^\mu \tau^I \Lelli \right) \left( H^\dagger \rmI \Dilr H \right)$ & $\Op^{\alpha\beta\gamma\lambda}_{\ell e}$ & $\left( \BLelli \gamma^\mu \Lelli \right) \left( \BREi[\gamma] \gamma^{}_\mu \REi[\lambda] \right) $
\\
\cline{1-2}\cline{5-6}
\multicolumn{2}{c|}{$\psi^2 X D$} & $\Op^{\prime(3)\alpha\beta}_{H\ell}$ & $ \left( \BLelli \rmI \Dilrs \Lelli \right) \left( H^\dagger \tau^I H \right)$ & \multicolumn{2}{c}{Semileptonic}
\\
\cline{1-2}\cline{5-6}
$\Op^{\alpha\beta}_{B\ell}$ & $\left( \BLelli \gamma^\mu \Lelli \right) D^\nu B^{}_{\mu\nu}$ & $\Op^{\alpha\beta}_{He}$ & $\left( \BREi \gamma^\mu \REi\right) \left( H^\dagger \rmI \Dlr H \right)$ & $\Op^{(1)\alpha\beta\gamma\lambda}_{\ell Q}$ & $ \left( \BLelli \gamma^\mu \Lelli \right) \left( \BLQi[\gamma] \gamma^{}_\mu \LQi[\lambda] \right)$
\\
$\Op^{\alpha\beta}_{W\ell}$ & $\left( \BLelli \gamma^\mu \tau^I \Lelli \right) \left( D^\nu W^{}_{\mu\nu} \right)^I$ & $\Op^{\prime\alpha\beta}_{He}$ & $ \left( \BREi \rmI \Dlrs \REi \right) \left( H^\dagger H \right) $ & $\Op^{(3)\alpha\beta\gamma\lambda}_{\ell Q}$ & $\left( \BLelli \gamma^\mu \tau^I \Lelli \right) \left( \BLQi[\gamma] \gamma^{}_\mu \tau^I \LQi[\lambda] \right)$
\\
& & & & $\Op^{\alpha\beta\gamma\lambda}_{\ell U}$ & $\left( \BLelli \gamma^\mu \Lelli \right) \left( \BRUi[\gamma] \gamma^{}_\mu \RUi[\lambda]\right)$

\\
& & & & $\Op^{\alpha\beta\gamma\lambda}_{\ell D}$ & $\left( \BLelli \gamma^\mu \Lelli \right) \left( \BRDi[\gamma] \gamma^{}_\mu \RDi[\lambda]\right)$
\\
\hline\hline
\end{tabular}}
\vspace{-0.15cm}
\caption{Dimension-six operators at the one-loop level induced by heavy Majorana neutrinos in the type-I seesaw model in the Green's basis, where the Hermitian conjugates of the operators in the classes $\psi^2 H D^2$, $\psi^2 X H$ and $\psi^2 H^3$, as well as those of the four-fermion operators, have not been listed explicitly.}
\label{tb:Gbasis}
\vspace{-0.25cm}
\end{table}

The dimension-six operators in the Green's basis induced by heavy Majorana neutrinos in the type-I seesaw at the one-loop level are listed in Table~\ref{tb:Gbasis} and the associated one-loop-level Wilson coefficients up to $\mathcal{O}\left( M^{-2} \right)$ are collected in the remaining part of this subsection, where an overall loop factor $1/\left( 4\pi\right)^2$ is not explicitly indicated in all one-loop-level Wilson coefficients. To make a distinction with the Wilson coefficients in the Warsaw basis, we shall denote the Wilson coefficients by ``$G$'', with different superscripts and subscripts, for the operators in the Green's basis.
\begin{itemize}
\item $X^2H^2$
\begin{eqnarray}
G^{}_{HB} &=& \frac{g^2_1}{24} {\rm tr} \left( \Yn M^{-2} \DYn \right) \;,
\label{eq:HB}
\\
G^{}_{HW} &=& \frac{g^2_2}{24} {\rm tr} \left( \Yn M^{-2} \DYn \right) \;,
\label{eq:HW}
\\
G^{}_{HWB} &=& \frac{g^{}_1 g^{}_2}{12} {\rm tr} \left( \Yn M^{-2} \DYn \right) \;.
\label{eq:HWB}
\end{eqnarray}

\item $H^2 X D^2$
\begin{eqnarray}
G^{}_{BDH} &=& -\frac{g^{}_1}{36} \DDYni{i} \pM{-2} \left( 5 + 6\lnmi \right) \;,
\label{eq:BDH}
\\
G^{}_{WDH} &=& -\frac{g^{}_2}{36} \DDYni{i} \pM{-2} \left( 5 + 6\lnmi \right) \;.
\label{eq:WDH}
\end{eqnarray}

\item $H^2 D^4$
\begin{eqnarray}
G^{}_{DH} = \frac{1}{3} {\rm tr} \left( \Yn M^{-2} \DYn \right) \;.
\label{eq:DH}
\end{eqnarray}

\item $H^4D^2$
\begin{eqnarray}
G^{}_{H\Box} &=& \frac{1}{2} \DDYni{k} \DDYni[k]{i} \frac{\pM{4} - 2\pM{2} \pM[k]{2} \lnik - \pM[k]{4}}{\left( \pM{2} - \pM[k]{2} \right)^3} - \frac{1}{4} \DDYni{k}\DDYni{k}
\nonumber
\\
&& \times \frac{\pM{6} - \left(7-2\lnik \right) \pM{4} \pM[k]{2} + \left( 7+2\lnik \right) \pM{2}\pM[k]{4} - \pM[k]{6} }{ \pM{}\pM[k]{} \left(\pM{2} - \pM[k]{2}\right)^3 } \;,
\label{eq:HBox}
\\
G^{}_{HD} &=& -\frac{1}{2} \left( \DYn \Yl \DYl \Yn \right)^{}_{ii} \pM{-2} \left(1 + 2\lnmi\right) - \frac{1}{2} \DDYni{k}\DDYni[k]{i} \frac{\lnik}{\pM{2}-\pM[k]{2}}
\nonumber
\\
&& - \DDYni{k}\DDYni{k} \frac{\left(1+\lnmk\right) \pM{2} - \left(1+\lnmi\right) \pM[k]{2}}{\pM{}\pM[k]{} \left( \pM{2} - \pM[k]{2}\right) } \;,
\label{eq:HD}
\\
G^\prime_{HD} &=& \frac{1}{2} \left( \DYn \Yl \DYl \Yn \right)^{}_{ii} \pM{-2} \left(1 + 2\lnmi\right) - \DDYni{k}\DDYni[k]{i} \frac{\lnik}{\pM{2}-\pM[k]{2}}
\nonumber
\\
&& - \frac{1}{2} \DDYni{k}\DDYni{k} \frac{\left(1+2\lnmk\right) \pM{2} - \left(1+2\lnmi\right) \pM[k]{2}}{\pM{}\pM[k]{} \left( \pM{2} - \pM[k]{2}\right) } \;,
\label{eq:HDP}
\\
G^{\prime\prime}_{HD} &=& -\frac{\rmI}{4} \DDYni{k}\DDYni{k} \frac{\pM{4} - 2\pM{2}\pM[k]{2}\lnik - \pM[k]{4}}{\pM{}\pM[k]{} \left(\pM{2}-\pM[k]{2}\right)^2} \;.
\label{eq:HDPP}
\end{eqnarray}

\item $H^6$
\begin{eqnarray}
G^{}_{H} &=& \frac{2}{3} \DDYni{k} \DDYni[k]{j} \DDYni[j]{i} \frac{\pM{2}\pM[j]{2}\lnji + \pM{2}\pM[k]{2}\lnik + \pM[j]{2}\pM[k]{2}\lnkj }{\left( \pM{2} - \pM[k]{2} \right) \left( \pM{2} - \pM[j]{2} \right) \left( \pM[k]{2} - \pM[j]{2} \right)}
\nonumber
\\
&& - 2  \pM{}\pM[j]{} \DDYni{k} \DDYni[k]{j} \DDYni{j} \frac{ \pM{2}\lnkj + \pM[j]{2}\lnik + \pM[k]{2}\lnji }{\left( \pM{2} - \pM[k]{2} \right) \left( \pM{2} - \pM[j]{2} \right) \left( \pM[k]{2} - \pM[j]{2} \right)} \;.
\label{eq:H}
\end{eqnarray}

\item $\psi^2 D^3$
\begin{eqnarray}
G^{\alpha\beta}_{\ell D} = - \frac{1}{3} \left( \Yn M^{-2} \DYn \right)^{}_{\alpha\beta} \;.
\label{eq:lD}
\end{eqnarray}

\item $\psi^2 X H$
\begin{eqnarray}
G^{\alpha\beta}_{eB} &=& \frac{g^{}_1}{8} \left( \Yn M^{-2} \DYn \Yl \right)^{}_{\alpha\beta} \;,
\label{eq:eB}
\\
G^{\alpha\beta}_{eW} &=& \frac{g^{}_2}{8} \left( \Yn M^{-2} \DYn \Yl \right)^{}_{\alpha\beta} \;.
\label{eq:eW}
\end{eqnarray}

\item $\psi^2HD^2$
\begin{eqnarray}
G^{\alpha\beta}_{eHD1} &=& \left( \Yn M^{-2} \DYn \Yl \right)^{}_{\alpha\beta} \;,
\label{eq:eHD1}
\\
G^{\alpha\beta}_{eHD2} &=& \frac{1}{4} \Yni{i} \pM{-2} \left( \DYn \Yl \right)^{}_{i\beta} \left(3+2\lnmi\right) \;,
\label{eq:eHD2}
\\
G^{\alpha\beta}_{eHD3} &=& \frac{1}{2} \left( \Yn M^{-2} \DYn \Yl \right)^{}_{\alpha\beta} \;,
\label{eq:eHD3}
\\
G^{\alpha\beta}_{eHD4} &=& \frac{1}{4} \Yni{i} \pM{-2} \left( \DYn \Yl \right)^{}_{i\beta} \left(7+2\lnmi\right) \;.
\label{eq:eHD4}
\end{eqnarray}

\item $\psi^2 X D$
\begin{eqnarray}
G^{\alpha\beta}_{B\ell} &=& \frac{g^{}_1}{72} \Yni{i} \pM{-2} \DYni{i} \left(11+6\lnmi\right) \;,
\label{eq:Bl}
\\
G^{\alpha\beta}_{W\ell} &=& -\frac{g^{}_2}{72} \Yni{i} \pM{-2} \DYni{i} \left(11+6\lnmi\right) \;.
\label{eq:Wl}
\end{eqnarray}

\item $\psi^2DH^2$
\begin{eqnarray}
G^{(1)\alpha\beta}_{H\ell} &=& \frac{g^2_1+3g^2_2}{32} \Yni{i} \pM{-2} \DYni{i} \left(11+6\lnmi\right) - \frac{1}{8} \Yni{i} \pM{-1} \left( Y^{\rm T}_\nu Y^\ast_\nu \right)^{}_{ik} \pM[k]{-1} \DYni{k}
\nonumber
\\
&& \times \left( 1 + \lnmi + \lnmk \right) - \frac{1}{2} \Yni{i} \left(\DYn\Yn\right)^{}_{ik} \DYni{k} \frac{\lnik}{\pM{2} - \pM[k]{2}}
\nonumber
\\
&& - \frac{1}{8} \Yni{i} \left( Y^{\rm T}_\nu Y^\ast_\nu \right)^{}_{ik} \DYni{k} \frac{\pM{2} \left(1+2\lnmk\right) - \pM[k]{2} \left(1+2\lnmi\right) }{\pM{}\pM[k]{}\left(\pM{2} - \pM[k]{2}\right)} \;,
\label{eq:Hl1}
\\
G^{\prime(1)\alpha\beta}_{H\ell} &=& \frac{3\lambda}{4} \Yni{i} \pM{-2} \DYni{i} \left(3+2\lnmi\right) - \frac{1}{8} \Yni{i} \pM{-1} \left( Y^{\rm T}_\nu Y^\ast_\nu\right)^{}_{ik} \pM[k]{-1} \DYni{k}
\nonumber
\\
&& \times \left( 3 + \lnmi + \lnmk \right) -\frac{1}{4} \Yni{i} \DDYni{k} \DYni{k} \frac{\lnik}{\pM{2} - \pM[k]{2}}
\nonumber
\\
&& -\frac{1}{8} \Yni{i} \left( Y^{\rm T}_\nu Y^\ast_\nu \right)^{}_{ik} \DYni{k} \frac{\pM{2} \left(3+2\lnmk\right) - \pM[k]{2} \left(3+2\lnmi\right) }{\pM{}\pM[k]{}\left(\pM{2} - \pM[k]{2}\right)} \;,
\label{eq:HlP1}
\\
G^{\prime\prime(1)\alpha\beta}_{H\ell} &=& \frac{\rmI}{4} \Yni{i} \DDYni{k} \DYni{k} \frac{\pM{2} \left(2-\lnik\right) - \pM[k]{2} \left(2+\lnik\right)}{\left(\pM{2} - \pM[k]{2}\right)^2}
\nonumber
\\
&& + \frac{\rmI}{4} \Yni{i} \left(Y^{\rm T}_\nu Y^\ast_\nu\right)^{}_{ik} \DYni{k} \frac{\pM{4} - 2\pM{2}\pM[k]{2}\lnik - \pM[k]{4} }{\pM{}\pM[k]{}\left( \pM{2} - \pM[k]{2} \right)^2} \;,
\label{eq:HlPP1}
\\
G^{(3)\alpha\beta}_{H\ell} &=& \frac{g^2_2-g^2_1}{32} \Yni{i} \pM{-2} \DYni{i} \left(11+6\lnmi\right) + \frac{1}{8} \Yni{i} \pM{-1} \left( Y^{\rm T}_\nu Y^\ast_\nu \right)^{}_{ik} \pM[k]{-1} \DYni{k}
\nonumber
\\
&& \times \left( 1 + \lnmi + \lnmk \right) \;,
\label{eq:Hl3}
\\
G^{\prime(3)\alpha\beta}_{H\ell} &=& -\frac{\lambda}{4} \Yni{i} \pM{-2} \DYni{i} \left(3+2\lnmi\right) + \frac{1}{8} \Yni{i} \pM{-1} \left( Y^{\rm T}_\nu Y^\ast_\nu \right)^{}_{ik} \pM[k]{-1} \DYni{k}
\nonumber
\\
&& \times \left( 3 + \lnmi + \lnmk \right) \;,
\label{eq:HlP3}
\\
G^{\alpha\beta}_{He} &=& \frac{1}{8} \left(\DYl \Yn\right)^{}_{\alpha i} \pM{-2} \left( \DYn \Yl \right)^{}_{i\beta} \left(1-2\lnmi\right) \;,
\label{eq:He}
\\
G^{\prime\alpha\beta}_{He} &=& \frac{1}{8} \left(\DYl \Yn\right)^{}_{\alpha i} \pM{-2} \left( \DYn \Yl \right)^{}_{i\beta} \left(3+2\lnmi\right) \;.
\label{eq:HeP}
\end{eqnarray}

\item $\psi^2 H^3$
\begin{eqnarray}
G^{\alpha\beta}_{eH} &=& - 2 \lambda \Yni{i} \pM{-2} \left(\DYn \Yl\right)^{}_{i\beta} \left(1+\lnmi\right) + \Yni{i} \DDYni{k} \left(\DYn \Yl\right)^{}_{k\beta} \frac{\lnik}{\pM{2} - \pM[k]{2}}
\nonumber
\\
&& + \Yni{i} \left(Y^{\rm T}_\nu Y^\ast_\nu\right)^{}_{ik} \left( \DYn \Yl \right)^{}_{k\beta} \frac{\pM{2}\left(1+\lnmk\right) - \pM[k]{2}\left(1+\lnmi\right)}{\pM{}\pM[k]{}\left(\pM{2} - \pM[k]{2} \right)} \;.
\label{eq:eH}
\end{eqnarray}

\item Four-lepton
\begin{eqnarray}
G^{\alpha\beta\gamma\lambda}_{\ell\ell} &=& -\frac{1}{8} \Yni{i} \pM{-2} \left[ \DYni{i} \left(\Yl\DYl\right)^{}_{\gamma\lambda} -\DYni[\lambda]{i} \left(\Yl\DYl\right)^{}_{\gamma\beta} \right] \left( 3+2\lnmi \right)
\nonumber
\\
&& -\frac{1}{8} \Yni{i} \DYni[\lambda]{i} \Yni[\gamma]{k}\DYni{k} \frac{\lnik}{\pM{2} - \pM[k]{2}} - \frac{1}{4} \Yni{k} \Yni[\gamma]{k} \DYni{i} \DYni[\lambda]{i}
\nonumber
\\
&& \times \frac{\pM{2}\left(1+\lnmk\right) - \pM[k]{2}\left(1+\lnmi\right)}{\pM{}\pM[k]{}\left(\pM{2} - \pM[k]{2}\right)} \;,
\label{eq:ll}
\\
G^{\alpha\beta\gamma\lambda}_{\ell e} &=& \frac{1}{8} \Yni{i} \pM{-2} \DYni{i} \left( \DYl\Yl\right)^{}_{\gamma\lambda} \left(3+2\lnmi\right) \;.
\label{eq:le}
\end{eqnarray}

\item Semileptonic
\begin{eqnarray}
G^{(1) \alpha\beta\gamma\lambda}_{\ell Q} &=& \frac{1}{16} \Yni{i} \pM{-2} \DYni{i} \left( \Yu\DYu  - \Yd\DYd \right)^{}_{\gamma\lambda} \left(3+2\lnmi\right) \;,
\label{eq:lQ1}
\\
G^{(3) \alpha\beta\gamma\lambda}_{\ell Q} &=& \frac{1}{16} \Yni{i} \pM{-2} \DYni{i} \left( \Yu\DYu  + \Yd\DYd \right)^{}_{\gamma\lambda} \left(3+2\lnmi\right) \;,
\label{eq:lQ3}
\\
G^{\alpha\beta\gamma\lambda}_{\ell U} &=& -\frac{1}{8} \Yni{i} \pM{-2} \DYni{i} \left( \DYu \Yu \right)^{}_{\gamma\lambda} \left(3+2\lnmi\right) \;,
\label{eq:lU}
\\
G^{\alpha\beta\gamma\lambda}_{\ell d} &=& \frac{1}{8} \Yni{i} \pM{-2} \DYni{i} \left( \DYd \Yd \right)^{}_{\gamma\lambda} \left(3+2\lnmi\right) \;.
\label{eq:lDD}
\end{eqnarray}

\end{itemize}

It is worth pointing out that the contributions from multiple heavy Majorana neutrinos in the loops to above Wilson coefficients may be symmetrized with respect to the generation index of heavy Majorana neutrinos. This is naturally expected, since heavy Majorana neutrinos appear as intermediate particles and must play an identical role when they are integrated out simultaneously at the matching scale $\mu \sim \mathcal{O} \left( M^{}_i \right)$. However, to maintain the results as concise as possible, we have not put the Wilson coefficients in a completely symmetric form. Additionally, the Wilson coefficient of $\Op^{\alpha\beta\gamma\lambda}_{\ell\ell}$, i.e., $G^{\alpha\beta\gamma\lambda}_{\ell\ell}$, can be symmetrized with respect to the former and latter two flavor indices due to the flavor structure of $\Op^{\alpha\beta\gamma\lambda}_{\ell\ell}$.

As mentioned before, these results in the Green's basis are not indispensable. One can directly convert the operators obtained by calculating the supertraces in Eq.~(\ref{eq:supertraces}) into the independent operators in the Warsaw basis by making use of algebraic, Fierz identities, integration by parts and the EOMs of relevant fields, but such a procedure is cumbersome and muddling. From this perspective, the above results in the Green's basis should be useful to make the simplification process smooth and the results trackable. In addition, it is convenient to compare the results with those derived by calculating Feynman diagrams, as an efficient cross-check. In the diagrammatic calculations the matching is usually done first with the operators in the Green's basis, which are then converted into those in the Warsaw basis, as in Refs.~\cite{Gherardi:2020det,Zhang:2021tsq}.

\section{Warsaw Basis}\label{sec:Warsaw}
To convert the operators in the Green's basis into those in the Warsaw basis, the EOMs of relevant fields are needed. From Eqs.~(\ref{eq:Lagrangian}) and (\ref{eq:Lagrangian-SM}), we can find the EOMs as follows
\begin{eqnarray}\label{eq:EOM}
\rmI \slashed{D} E^{}_{\rm R} &=& H^\dagger Y^\dagger_l \ell^{}_{\rm L} \;,
\nonumber
\\
\rmI \slashed{D} \ell^{}_{\rm L} &=& Y^{}_l H E^{}_{\rm R} +  Y^{}_\nu \widetilde{H} P^{}_{\rm R} N \;,
\nonumber
\\
D^\nu B^{}_{\mu \nu} &=& \frac{1}{2} g^{}_1 \left[ H^\dagger \rmI \Dlr H + 2 \sum^{}_f Y(f) \overline{f} \gamma^{}_\mu f \right] \;,
\nonumber
\\
\left( D^\nu W_{\mu \nu} \right)^I &=& \frac{1}{2} g^{}_2 \left( H^\dagger \rmI \Dilr H + \overline{Q^{}_{\rm L}} \tau^I \gamma^{}_\mu Q^{}_{\rm L} + \overline{\ell^{}_{\rm L}} \tau^I \gamma^{}_\mu \ell^{}_{\rm L} \right) \;,
\nonumber
\\
\left( D^2 H \right)^a &=& - m^2 H^a - 2\lambda \left( H^\dagger H \right) H^a - \overline{E^{}_{\rm R}} Y^\dagger_l \ell^a_{\rm L} - \overline{D^{}_{\rm R}} Y^\dagger_{\rm d} Q^a_{\rm L} + \epsilon^{ab} \overline{Q^b_{\rm L}} Y^{}_{\rm u}  U^{}_{\rm R} + \epsilon^{ab} \overline{\ell^b_{\rm L}} Y^{}_\nu P^{}_{\rm R} N \;,\;~~
\end{eqnarray}
where $Y(f)$ is the hypercharge for $f= Q^{}_{\rm L}, U^{}_{\rm R}, D^{}_{\rm R}, \ell^{}_{\rm L}, E^{}_{\rm R}$, and $I =1,2,3$ and $a,b=1,2$. Again, the heavy fields $N$ in Eq.~(\ref{eq:EOM}) should be replaced with the terms on the right-hand side of Eq.~(\ref{eq:classical-N}).

\subsection{Threshold corrections}
Applying the EOMs in Eq.~(\ref{eq:EOM}) to the dimension-six operators in the Green's basis, one obtains one-loop corrections to the renormalizable terms in the Warsaw basis:
\begin{eqnarray}\label{eq:GtoW}
\delta Z^{}_H &=& \delta Z^{\rm G}_H \;,
\nonumber
\\
\delta Z^{}_\ell &=& \delta Z^{\rm G}_\ell \;,
\nonumber
\\
\delta m^2 &=& \left( \delta m^2 \right)^{\rm G} + m^4 G^{}_{DH} \;,
\nonumber
\\
\delta \lambda &=& \delta \lambda^{\rm G} + 2g^{}_2 m^2 G^{}_{WDH} + 4\lambda m^2 G^{}_{DH} + m^2 G^\prime_{HD} \;,
\nonumber
\\
\left( \delta \Yl \right)^{}_{\alpha\beta} &=& \left( \delta Y^{\rm G}_\ell \right)^{}_{\alpha\beta} + m^2 \left[ \Yli{\beta} G^{}_{DH} - G^{\alpha\beta}_{eHD1} - \frac{1}{2} G^{\alpha\beta}_{eHD2} + \frac{1}{2} G^{\alpha\beta}_{eHD4} \right] \;,
\nonumber
\\
\left( \delta \Yu \right)^{}_{\alpha\beta} &=& m^2 \Yui{\beta} G^{}_{DH} \;,
\nonumber
\\
\left( \delta \Yd \right)^{}_{\alpha\beta} &=& m^2 \Ydi{\beta} G^{}_{DH} \;.
\end{eqnarray}
One can see that $\delta Y^{}_{\rm u}$ and $\delta Y^{}_{\rm d}$ absent in the Green's basis now appear in the Warsaw basis. They are induced by the operator $\mathcal{O}^{}_{DH}$ in the Green's basis after applying the EOM of $H$. $\Op^{}_{DH}$ also gives an additional one-loop contribution to the Wilson coefficient of the dimension-five operator
\begin{eqnarray}
\left( \delta C^{(5)} \right)^{}_{\alpha\beta} &=& \left( \delta C^{(5)} \right)^{\rm G}_{\alpha\beta} - 2\mu^2 C^{(5)}_{\alpha\beta} G^{}_{DH} \;,
\end{eqnarray}
but this contribution is of the order of $\mathcal{O}\left(M^{-3} \right)$ and will be omitted. Then the kinetic terms of $H$ and $\ell$ need to be normalized via
\begin{eqnarray}\label{eq:shift}
H \to \left( 1 - \frac{1}{2} \delta Z^{}_H \right) H \;,\quad \ell \to \left( 1 - \frac{1}{2} \delta Z^{}_\ell \right) \ell \;,
\end{eqnarray}
where only the leading-order terms are kept. These redefinitions of fields give additional one-loop contributions to the effective couplings in the EFT via the tree-level terms. From Eqs.~(\ref{eq:GtoW})-(\ref{eq:shift}) and the results in Sec.~\ref{sec:Green}, we obtain the effective couplings in the EFT
\begin{eqnarray}
m^2_{\rm eff} &=& m^2 \left( 1 - \delta Z^{}_H \right) - \delta m^2
\nonumber
\\
&=& m^2 - \frac{1}{\left(4\pi\right)^2} \DDYni{i} \left[ \frac{m^2}{2}  \left( 1 + 2 \lnmi \right) + \frac{m^4}{3 \pM{2}} - 2 M^2_i \left( 1+\lnmi \right) \right]  \;,
\nonumber
\\
\lambda^{}_{\rm eff} &=& \lambda \left( 1 - 2\delta Z^{}_H \right) - \delta \lambda
\nonumber
\\
&=& \lambda + \frac{1}{\left(4\pi\right)^2} \left\{ \DDYni{i} \left[ - \lambda \left( 1 + 2 \lnmi \right) - \frac{4\lambda m^2}{3\pM{2}} + \frac{g^2_2 m^2}{18 \pM{2}} \left(5+6\lnmi\right) \right]  \right.
\nonumber
\\
&& - \frac{m^2}{2 \pM{2}} \left( \DYn \Yl \DYl \Yn \right)^{}_{ii} \left(1+2\lnmi\right) + \left[m^2\lnik + \pM{2}\left( 1+\lnmi \right) - \pM[k]{2}\left( 1+\lnmk \right) \right]
\nonumber
\\
&& \times \frac{\DDYni{k} \DDYni[k]{i}}{\pM{2} - \pM[k]{2}} + \left[ m^2 \pM{2} \left( 1+2\lnmk \right) - m^2 \pM[k]{2} \left( 1+ 2\lnmi\right) - 2 \pM{2} \pM[k]{2} \lnik \right]
\nonumber
\\
&& \times \left. \frac{\DDYni{k} \DDYni{k}}{2\pM{}\pM[k]{}\left( \pM{2} - \pM[k]{2} \right) }  \right\} \;,
\nonumber
\\
\left(Y^{\rm eff}_l \right)^{}_{\alpha\beta} &=& \left[ \Yl \left( 1 - \frac{1}{2} \delta Z^{}_H \right) - \frac{1}{2} \delta Z^{}_\ell \Yl - \delta \Yl \right]^{}_{\alpha\beta}
\nonumber
\\
&=& \Yli{\beta} - \frac{1}{\left(4\pi\right)^2} \left\{ \Yli{\beta} \DDYni{i} \left[ \frac{1}{4} \left( 1 + 2 \lnmi \right) + \frac{m^2}{3 \pM{2}} \right] - \frac{1}{8} \Yni{i} \left( \DYn\Yl \right)^{}_{i\beta} \right.
\nonumber
\\
&& \times \left. \left[5 + \frac{6m^2}{M^2_i} + 2 \left( 3 + \frac{2m^2}{M^2_i} \right)\lnmi \right] \right\} \;,
\nonumber
\\
\left( Y^{\rm eff}_{\rm u} \right)^{}_{\alpha\beta} &=&  \left[ \Yu \left( 1 - \frac{1}{2} \delta Z^{}_H \right) - \delta \Yu \right]^{}_{\alpha\beta}
\nonumber
\\
&=& \Yui{\beta} - \frac{1}{\left(4\pi\right)^2} \Yui{\beta} \DDYni{i} \left[ \frac{1}{4} \left( 1 + 2 \lnmi \right) + \frac{m^2}{3 \pM{2}} \right] \;,
\nonumber
\\
\left( Y^{\rm eff}_{\rm d} \right)^{}_{\alpha\beta} &=&  \left[ \Yd \left( 1 - \frac{1}{2} \delta Z^{}_H \right) - \delta \Yd \right]^{}_{\alpha\beta}
\nonumber
\\
&=& \Ydi{\beta} - \frac{1}{\left(4\pi\right)^2} \Ydi{\beta} \DDYni{i} \left[ \frac{1}{4} \left( 1 + 2 \lnmi \right) + \frac{m^2}{3\pM{2}} \right]  \;,
\end{eqnarray}
up to $\mathcal{O}\left( M^{-2} \right)$. Similarly, the Wilson coefficient of the dimension-five operator is given by
\begin{eqnarray}
\left( C^{(5)}_{\rm eff} \right)^{}_{\alpha\beta} &=& \left[ C^{(5)} \left( 1 - \delta Z^{}_H \right) - \frac{1}{2} \delta Z^{}_\ell C^{(5)} - \frac{1}{2} C^{(5)} \delta Z^{\rm T}_\ell + \delta C^{(5)} \right]^{}_{\alpha\beta}
\nonumber
\\
&=& C^{(5)}_{\alpha\beta} - \frac{1}{\left(4\pi\right)^2} \left\{ \frac{1}{2} C^{(5)}_{\alpha\beta} \DDYni{i} \left( 1 + 2 \lnmi \right) + \frac{1}{8} \Yni{i} \left( \DYn C^{(5)} \right)^{}_{i\beta} \left( 3+2\lnmi \right) \right.
\nonumber
\\
&& + \frac{1}{8} C^{(5)}_{\alpha\gamma} \Yni[\beta]{i} \DYni[\gamma]{i} \left( 3+2\lnmi \right) - \left[ 2\lambda \left( 1+\lnmi \right) + \frac{g^2_1+g^2_2}{4} \left( 1+3\lnmi\right) \right]
\nonumber
\\
&& \times \left. \Yni{i} \pM{-1} \left( Y^{\rm T}_\nu \right)^{}_{i\beta} \phantom{\frac{1}{1}}\hspace{-4mm} \right\} \;,
\end{eqnarray}
up to $\mathcal{O}\left( M^{-2} \right)$. These are the complete one-loop matching results, which can be used together with the two-loop RGEs of relevant Wilson coefficients in the SEFT.

\subsection{Dimension-six operators}

In the type-I seesaw model, $\mathcal{O}^{(1)\alpha\beta}_{H\ell}$ and $\mathcal{O}^{(3)\alpha\beta}_{H\ell}$ can already be obtained from the tree-level matching as shown in Eq.~(\ref{eq:6-tree}). Their Wilson coefficients at the tree level are given by
\begin{eqnarray}\label{eq:6-tree-WC}
\left[ C^{(1)}_{H\ell} \right]^{\alpha\beta}_{\rm tree} = - \left[ C^{(3)}_{H\ell} \right]^{\alpha\beta}_{\rm tree} = \frac{1}{4} C^{(6)}_{\alpha\beta} = \frac{1}{4} \left( Y^{}_\nu M^{-2} Y^\dagger_\nu \right)^{}_{\alpha\beta} \;.
\end{eqnarray}
These tree-level Wilson coefficients result in extra one-loop contributions to the total Wilson coefficients of $\mathcal{O}^{(1)\alpha\beta}_{H\ell}$ and $\mathcal{O}^{(3)\alpha\beta}_{H\ell}$ via the normalizations of the kinetic terms of $H$ and $\ell$ given in Eq.~(\ref{eq:shift}), i.e.,
\begin{eqnarray}
\delta C^{(1)\alpha\beta}_{H\ell} &=& - \delta C^{(3)\alpha\beta}_{H\ell} = - \frac{1}{4} \left( C^{(6)} \delta Z^{}_H  + \frac{1}{2} \delta Z^{}_\ell C^{(6)} + \frac{1}{2} C^{(6)} \delta Z^{}_\ell \right)^{}_{\alpha\beta}
\nonumber
\\
&=& - \frac{1}{8 \left(4\pi\right)^2} \left[ \left( \Yn M^{-2} \DYn \right)^{}_{\alpha\beta} \DDYni{i} \left( 1 + 2 \lnmi \right) + \frac{1}{4} \Yni{i} \left( \DYn \Yn M^{-2} \DYn \right)^{}_{i\beta} \right.
\nonumber
\\
&&  \times \left.  \left( 3+2\lnmi \right) + \frac{1}{4} \left( \Yn M^{-2} \DYn \Yn \right)^{}_{\alpha i} \DYni{i}  \left( 3+2\lnmi\right) \right]  \;,
\end{eqnarray}
which will be added into the total one-loop-level Wilson coefficients of $\mathcal{O}^{(1)\alpha\beta}_{H\ell}$ and $\mathcal{O}^{(3)\alpha\beta}_{H\ell}$.

All dimension-six operators in the Warsaw basis induced by the type-I seesaw model at the one-loop level are listed in Table~\ref{tb:Wbasis} and the associated one-loop-level Wilson coefficients up to $\mathcal{O} \left( M^{-2} \right)$ are explicitly given in the remaining part of this subsection, where an overall loop factor $1/\left( 4\pi\right)^2$ is implied in all Wilson coefficients.

\begin{table}[t]
\centering
\renewcommand\arraystretch{1.5}
\resizebox{\textwidth}{!}{
\begin{tabular}{c|c|c|c|c|c}
\hline\hline
\multicolumn{2}{c}{$X^2H^2$} & \multicolumn{2}{|c|}{$\psi^2DH^2$} & \multicolumn{2}{c}{Four-quark}
\\
\hline
$\Op^{}_{HB}$ & $B^{}_{\mu\nu} B^{\mu\nu} H^\dagger H$ & $\Op^{(1)\alpha\beta}_{HQ}$ &  $\left( \BLQi \gamma^\mu \LQi \right) \left( H^\dagger \rmI \Dlr H \right)$ & $\Op^{(1)\alpha\beta\gamma\lambda}_{QU}$ & $\left( \BLQi \gamma^\mu \LQi \right) \left( \BRUi[\gamma] \gamma^{}_\mu \RUi[\lambda] \right)$
\\
$\Op^{}_{HW}$ & $W^I_{\mu\nu} W^{I\mu\nu} H^\dagger H$ & $\Op^{(3)\alpha\beta}_{HQ} $ & $ \left( \BLQi \gamma^\mu \tau^I \LQi \right) \left(H^\dagger \rmI \Dilr H \right)$ & $\Op^{(8)\alpha\beta\gamma\lambda}_{QU}$ & $\left(\BLQi \gamma^\mu T^A \LQi \right) \left( \BRUi[\gamma] \gamma^{}_\mu T^A \RUi[\lambda] \right)$
\\
$\Op^{}_{HWB}$ & $W^I_{\mu\nu} B^{\mu\nu} \left( H^\dagger \tau^I H \right)$ & $\Op^{\alpha\beta}_{HU}$ & $\left( \BRUi \gamma^\mu \RUi \right) \left(H^\dagger \rmI \Dlr H \right)$ & $ \Op^{(1)\alpha\beta\gamma\lambda}_{Qd}$ & $ \left( \BLQi \gamma^\mu \LQi \right) \left( \BRDi[\gamma] \gamma^{}_\mu \RDi[\lambda] \right)$
\\
\cline{1-2}
\multicolumn{2}{c|}{$H^4D^2$} & $\Op^{\alpha\beta}_{Hd}$ & $\left( \BRDi \gamma^\mu \RDi \right) \left( H^\dagger \rmI \Dlr H \right)$ & $\Op^{(8)\alpha\beta\gamma\lambda}_{Qd}$ & $\left( \BLQi \gamma^\mu T^A \LQi \right) \left( \BRDi[\gamma] \gamma^{}_\mu T^A \RDi[\lambda] \right)$
\\
\cline{1-2}
$\Op^{}_{H\Box}$ & $\left(H^\dagger H\right) \Box \left( H^\dagger H \right)$ & $\Op^{(1)\alpha\beta}_{H\ell}$ & $\left( \BLelli \gamma^\mu \Lelli \right) \left( H^\dagger \rmI \Dlr H \right)$ & $\Op^{(1)\alpha\beta\gamma\lambda}_{QUQd}$ & $ \left( \overline{Q^a_{\alpha{\rm L}}} \RUi \right) \epsilon^{ab} \left( \overline{Q^b_{\gamma{\rm L}}} \RDi[\lambda] \right) $
\\
\cline{5-6}
$\Op^{}_{HD}$ & $\left(H^\dagger D^{}_\mu H \right)^\ast \left(H^\dagger D^\mu H \right)$ & $\Op^{(3)\alpha\beta}_{H\ell}$ & $ \left( \BLelli \gamma^\mu \tau^I \Lelli \right) \left( H^\dagger \rmI \Dilr H \right) $ & \multicolumn{2}{c}{Four-lepton}
\\
\cline{1-2}\cline{5-6}
\multicolumn{2}{c|}{$H^6$} & $\Op^{\alpha\beta}_{He}$ & $\left(\BREi \gamma^\mu \REi \right) \left( H^\dagger \rmI \Dlr H \right)$ & $\Op^{\alpha\beta\gamma\beta}_{\ell\ell}$ & $ \left( \BLelli \gamma^\mu \Lelli \right) \left( \BLelli[\gamma] \gamma^{}_\mu \Lelli[\lambda] \right) $
\\
\cline{1-4}
$\Op^{}_H$ & $\left(H^\dagger H \right)^3$ & \multicolumn{2}{c|}{$\psi^2 H^3$} & $\Op^{\alpha\beta\gamma\lambda}_{\ell e}$ & $\left(\BLelli \gamma^\mu \Lelli \right) \left( \BREi[\gamma] \gamma^{}_\mu \REi[\lambda] \right) $
\\
\cline{1-4}
\multicolumn{2}{c|}{$\psi^2XH$} & $\Op^{\alpha\beta}_{UH}$ & $\left( \BLQi \widetilde{H} \RUi \right) \left( H^\dagger H \right) $ & &
\\
\cline{1-2}
$\Op^{\alpha\beta}_{eB}$ & $\left( \BLelli \sigma^{\mu\nu} \REi \right) H B^{}_{\mu\nu}$ & $\Op^{\alpha\beta}_{dH}$ & $ \left( \BLQi H \RDi \right) \left( H^\dagger H \right) $ & &
\\
$\Op^{\alpha\beta}_{eW}$ & $ \left( \BLelli \sigma^{\mu\nu} \REi \right) \tau^I H W^I_{\mu\nu} $ & $\Op^{\alpha\beta}_{eH}$ & $ \left( \BLelli H \REi \right) \left( H^\dagger H \right)$ & &
\\
\hline
\multicolumn{6}{c}{Semi-leptonic}
\\
\hline
$\Op^{(1)\alpha\beta\gamma\lambda}_{\ell Q}$ & $ \left( \BLelli \gamma^\mu \Lelli \right) \left( \BLQi[\gamma] \gamma^{}_\mu \LQi[\lambda] \right)$ & $\Op^{\alpha\beta\gamma\lambda}_{\ell U}$ & $\left( \BLelli \gamma^\mu \Lelli \right) \left( \BRUi[\gamma] \gamma^{}_\mu \RUi[\lambda] \right)$ & $\Op^{\alpha\beta\gamma\lambda}_{\ell e d Q}$ & $ \left( \BLelli \REi \right) \left( \BRDi[\gamma] \LQi[\lambda] \right) $
\\
$\Op^{(3)\alpha\beta\gamma\lambda}_{\ell Q}$ & $ \left( \BLelli \gamma^\mu \tau^I \Lelli \right) \left( \BLQi[\gamma] \gamma^{}_\mu \tau^I \LQi[\lambda] \right) $ & $\Op^{\alpha\beta\gamma\lambda}_{\ell d}$ & $\left( \BLelli \gamma^\mu \Lelli \right) \left( \RDi[\gamma] \gamma^{}_\mu \RDi[\lambda] \right)$ & $\Op^{(1)\alpha\beta\gamma\lambda}_{\ell e Q U}$ & $\left( \overline{\ell^a_{\alpha \rm L}} \REi \right) \epsilon^{ab} \left( \overline{Q^b_{\gamma \rm L}} \RUi[\lambda] \right)$
\\
\hline\hline
\end{tabular}}
\vspace{-0.15cm}
\caption{Dimension-six operators induced by the type-I seesaw model at the one-loop level in the Warsaw basis, where the Hermitian conjugates of the operators in classes $\psi^2 X H$ and $\psi^2 H^3$, as well as those of the four-fermion operators, have not been listed explicitly.}
\label{tb:Wbasis}
\vspace{-0.25cm}
\end{table}

\begin{itemize}
\item $X^2H^2$
\begin{eqnarray}
C^{}_{HB} &=& G^{}_{HB} = \frac{g^2_1}{24} {\rm tr} \left( \Yn M^{-2} \DYn \right) \;,
\label{eq:HBw}
\\
C^{}_{HWB} &=& G^{}_{HWB} = \frac{g^{}_1 g^{}_2}{12} {\rm tr} \left( \Yn M^{-2} \DYn \right) \;,
\label{eq:HWBw}
\\
C^{}_{HW} &=& G^{}_{HW} = \frac{g^2_2}{24} {\rm tr} \left( \Yn M^{-2} \DYn \right) \;.
\label{eq:HWw}
\end{eqnarray}

\item $H^4D^2$
\begin{eqnarray}
C^{}_{H\Box} &=& G^{}_{H\Box} + \frac{3g^{}_2}{2} G^{}_{WDH} + \frac{g^{}_1}{2} G^{}_{BDH} + \frac{1}{2} G^\prime_{HD}
\nonumber
\\
&=& -\frac{g^2_1+3g^2_2}{72} \DDYni{i} \pM{-2} \left(5+6\lnmi\right) + \frac{1}{4} \left( \DYn \Yl \DYl \Yn \right)^{}_{ii} \pM{-2} \left(1+2\lnmi\right)
\nonumber
\\
&& + \frac{1}{2} \DDYni{k} \DDYni[k]{i} \frac{\pM{4}\left(1-\lnik\right) - \pM[k]{4}\left(1+\lnik\right)}{\left( \pM{2} - \pM[k]{2} \right)^3} - \frac{\DDYni{k}\DDYni{k}}{2\pM{}\pM[k]{}\left( \pM{2} - \pM[k]{2} \right)^3}
\nonumber
\\
&& \times \left[ \pM{6} \left(1+\lnmk\right) - \pM{4}\pM[k]{2} \left(5+2\lnmi+\lnmk\right) + \pM{2}\pM[k]{4} \left( 5+\lnmi+2\lnmk \right) \right.
\nonumber
\\
&& - \left. \pM[k]{6} \left( 1+\lnmi \right)  \right] \;,\qquad
\label{eq:HBoxw}
\\
C^{}_{HD} &=& G^{}_{HD} + 2g^{}_1 G^{}_{BDH}
\nonumber
\\
&=& -\frac{g^2_1}{18} \DDYni{i} \pM{-2} \left(5+6\lnmi\right) - \frac{1}{2} \left(\DYn \Yl \DYl \Yn\right)^{}_{ii} \pM{-2} \left(1+2\lnmi\right) - \frac{\lnik}{2\left(\pM{2} - \pM[k]{2}\right)}
\nonumber
\\
&& \times \DDYni{k} \DDYni[k]{i} - \DDYni{k}\DDYni{k} \frac{\pM{2} \left( 1+\lnmk\right) - \pM[k]{2} \left(1+\lnmi\right) }{\pM{}\pM[k]{} \left(\pM{2} - \pM[k]{2} \right)} \;.
\label{eq:HDw}
\end{eqnarray}

\item $H^6$
\begin{eqnarray}
C^{}_H &=& G^{}_H + 4g^{}_2 \lambda G^{}_{WDH} + 4\lambda^2 G^{}_{DH} + 2\lambda G^\prime_{HD}
\nonumber
\\
&=& \frac{\lambda}{3 \pM{2}} \DDYni{i} \left[ 4\lambda  - \frac{g^2_2}{3} \left(5+6\lnmi\right) \right] + \lambda \left( \DYn \Yl \DYl \Yn \right)^{}_{ii} \pM{-2} \left(1+2\lnmi\right)
\nonumber
\\
&&  - 2\lambda \DDYni{k} \DDYni[k]{i} \frac{\lnik}{\pM{2} - \pM[k]{2}} - \lambda \DDYni{k} \DDYni{k}
\nonumber
\\
&& \times \frac{\pM{2} \left(1+2\lnmk\right) - \pM[k]{2} \left(1+2\lnmi\right) }{\pM{}\pM[k]{}\left( \pM{2} - \pM[k]{2} \right)} + \frac{2}{3} \DDYni{k}\DDYni[k]{j} \DDYni[j]{i}
\nonumber
\\
&& \times \frac{\pM{2}\pM[j]{2}\lnji + \pM{2}\pM[k]{2}\lnik + \pM[j]{2}\pM[k]{2}\lnkj}{\left(\pM{2} - \pM[k]{2}\right) \left(\pM{2} -\pM[j]{2}\right) \left(\pM[k]{2} - \pM[j]{2}\right)} - 2 \pM{}\pM[j]{} \DDYni{k} \DDYni[k]{j} \DDYni{j}
\nonumber
\\
&& \times \frac{\pM{2}\lnkj + \pM[k]{2}\lnji + \pM[j]{2}\lnik}{\left(\pM{2} - \pM[k]{2}\right) \left(\pM{2} - \pM[j]{2}\right) \left(\pM[k]{2} - \pM[j]{2}\right) } \;.
\label{eq:Hw}
\end{eqnarray}

\item $\psi^2 X H$
\begin{eqnarray}
C^{\alpha\beta}_{eB} &=& G^{\alpha\beta}_{eB} - \frac{g^{}_1}{8} \Yli[\gamma]{\beta} G^{\alpha\gamma}_{\ell D} - \frac{g^{}_1}{8} G^{\alpha\beta}_{eHD2} - \frac{g^{}_1}{2} G^{\alpha\beta}_{eHD3} + \frac{g^{}_1}{8} G^{\alpha\beta}_{eHD4}
\nonumber
\\
&=& \frac{g^{}_1}{24} \left( \Yn M^{-2} \DYn \Yl \right)^{}_{\alpha\beta} \;,
\label{eq:eBw}
%\end{eqnarray}
%\begin{eqnarray}
\\
C^{\alpha\beta}_{eW} &=& G^{\alpha\beta}_{eW} + \frac{g^{}_2}{8} \Yli[\gamma]{\beta} G^{\alpha\gamma}_{\ell D} - \frac{g^{}_2}{8} G^{\alpha\beta}_{eHD2} + \frac{g^{}_2}{8} G^{\alpha\beta}_{eHD4}
\nonumber
\\*
&=& \frac{5g^{}_2}{24} \left( \Yn M^{-2} \DYn \Yl \right)^{}_{\alpha\beta} \;.
\label{eq:eWw}
\end{eqnarray}

\item $\psi^2 D H^2$
\begin{eqnarray}
C^{(1)\alpha\beta}_{HQ} &=& \frac{g^{}_1}{6} \delta^{\alpha\beta} G^{}_{BDH} = - \frac{g^2_1}{216} \delta^{\alpha\beta} \DDYni{i} \pM{-2} \left(5+6\lnmi\right) \;,
\label{eq:HQ1w}
\\
C^{(3)\alpha\beta}_{HQ} &=& \frac{g^{}_2}{2} \delta^{\alpha\beta} G^{}_{WDH} = -\frac{g^2_2}{72} \delta^{\alpha\beta} \DDYni{i} \pM{-2} \left(5+6\lnmi\right) \;,
\label{eq:HQ3w}
\\
C^{\alpha\beta}_{HU} &=& \frac{2g^{}_1}{3} \delta^{\alpha\beta} G^{}_{BDH} = - \frac{g^2_1}{54} \delta^{\alpha\beta} \DDYni{i} \pM{-2} \left(5+6\lnmi\right) \;,
\label{eq:HUw}
\\
C^{\alpha\beta}_{Hd} &=& -\frac{g^{}_1}{3} \delta^{\alpha\beta} G^{}_{BDH} = \frac{g^2_1}{108} \delta^{\alpha\beta} \DDYni{i} \pM{-2} \left(5+6\lnmi\right) \;,
\label{eq:HDDw}
\end{eqnarray}
\vspace{-0.8cm}
\begin{eqnarray}
C^{(1)\alpha\beta}_{H\ell} &=& G^{(1)\alpha\beta}_{H\ell} - \frac{g^{}_1}{2} \delta^{\alpha\beta} G^{}_{BDH} + \frac{g^{}_1}{2} G^{\alpha\beta}_{B\ell} + \delta C^{(1)\alpha\beta}_{H\ell} - \frac{1}{8} \left[ \DYli[\gamma]{\beta} G^{\alpha\gamma}_{eHD2} + \Yli{\gamma} G^{\ast\beta\gamma}_{eHD2} \right]
\nonumber
\\
&& + \frac{1}{4} \left[ \DYli[\gamma]{\beta} G^{\alpha\gamma}_{eHD3} + \Yli{\gamma} G^{\ast\beta\gamma}_{eHD3} \right] - \frac{1}{8} \left[ \DYli[\gamma]{\beta} G^{\alpha\gamma}_{eHD4} + \Yli{\gamma} G^{\ast\beta\gamma}_{eHD4} \right]
\nonumber
\\
&=& \Yni{i} \pM{-2} \DYni{i} \left[ -\frac{1}{8} \DDYni[k]{k} \left( 1 + 2\lnmk \right) + \frac{11g^2_1+27g^2_2}{288} \left(11+6\lnmi\right) \right]
\nonumber
\\
&&  - \frac{1}{32} \left( 3+2\lnmi \right) \left[ \Yni{i} \left( \DYn \Yn M^{-2} \DYn \right)^{}_{i\beta} +  \left( \Yn M^{-2} \DYn \Yn \right)^{}_{\alpha i} \DYni{i} \right]
\nonumber
\\
&&  - \frac{1}{16} \left(3+2\lnmi\right) \left[ \Yni{i} \pM{-2} \left(\DYn \Yl \DYl\right)^{}_{i\beta} + \left( \Yl \DYl \Yn \right)^{}_{\alpha i} \pM{-2} \DYni{i}  \right]
\nonumber
\\
&&+ \frac{g^2_1}{72} \delta^{\alpha\beta} \DDYni{i} \pM{-2} \left(5+6\lnmi\right) - \frac{1}{8} \Yni{i} \pM{-1} \left(Y^{\rm T}_\nu Y^\ast_\nu\right)^{}_{ik} \pM[k]{-1} \DYni{k}
\nonumber
\\
&& \times \left(1+\lnmi+\lnmk\right)  - \frac{1}{2} \Yni{i} \DDYni{k} \DYni{k} \frac{\lnik}{\pM{2} - \pM[k]{2}}
\nonumber
\\
&& - \frac{1}{8} \Yni{i} \left( Y^{\rm T}_\nu Y^\ast_\nu \right)^{}_{ik} \DYni{k} \frac{\pM{2}\left(1+2\lnmk\right) - \pM[k]{2}\left(1+2\lnmi\right) }{\pM{}\pM[k]{} \left(\pM{2} - \pM[k]{2}\right)} \;,
\label{eq:Hl1w}
%\end{eqnarray}
%\vspace{-0.8cm}
%\begin{eqnarray}
\\
C^{(3)\alpha\beta}_{H\ell} &=& G^{(3)\alpha\beta}_{H\ell} + \frac{g^{}_2}{2} \delta^{\alpha\beta} G^{}_{WDH} + \frac{g^{}_2}{2} G^{\alpha\beta}_{W\ell} + \delta C^{(3)\alpha\beta}_{H\ell} - \frac{1}{8} \left[ \DYli[\gamma]{\beta} G^{\alpha\gamma}_{eHD2} + \Yli{\gamma} G^{\ast\beta\gamma}_{eHD2} \right]
\nonumber
\\
&& + \frac{1}{4} \left[ \DYli[\gamma]{\beta} G^{\alpha\gamma}_{eHD3} + \Yli{\gamma} G^{\ast\beta\gamma}_{eHD3} \right] - \frac{1}{8} \left[ \DYli[\gamma]{\beta} G^{\alpha\gamma}_{eHD4} + \Yli{\gamma} G^{\ast\beta\gamma}_{eHD4} \right]
\nonumber
\\
&=& \Yni{i} \pM{-2} \DYni{i} \left[  \frac{1}{8} \DDYni[k]{k} \left( 1 + 2\lnmk \right) + \frac{7g^2_2-9g^2_1}{288}  \left(11+6\lnmi\right) \right]
\nonumber
\\
&& + \frac{1}{32} \left( 3+2\lnmi \right) \left[ \Yni{i} \left( \DYn \Yn M^{-2} \DYn \right)^{}_{i\beta}  + \left( \Yn M^{-2} \DYn \Yn \right)^{}_{\alpha i} \DYni{i} \right]
\nonumber
\\
&&  - \frac{1}{16} \left(3+2\lnmi\right) \left[ \Yni{i} \pM{-2} \left( \DYn \Yl \DYl \right)^{}_{i\beta} + \left(\Yl \DYl \Yn \right)^{}_{\alpha i} \pM{-2} \DYni{i} \right]
\nonumber
\\
&& -\frac{g^2_2}{72} \delta^{\alpha\beta} \DDYni{i} \pM{-2} \left(5+6\lnmi\right)  + \frac{1}{8} \Yni{i} \pM{-1} \left(Y^{\rm T}_\nu Y^\ast_\nu \right)^{}_{ik} \pM[k]{-1} \DYni{k}
\nonumber
\\
&& \times \left(1+\lnmi+\lnmk\right) \;,
\label{eq:Hl3w}
\\
C^{\alpha\beta}_{He} &=& G^{\alpha\beta}_{He} - g^{}_1 \delta^{\alpha\beta} G^{}_{BDH} - \frac{1}{2} \DYli[\alpha]{\gamma} \Yli[\lambda]{\beta} G^{\gamma\lambda}_{\ell D} + \frac{1}{4} \left[ \DYli[\alpha]{\gamma} G^{\gamma\beta}_{eHD2} + \Yli[\gamma]{\beta} G^{\ast\gamma\alpha}_{eHD2} \right]
\nonumber
\\
&& - \frac{1}{4} \left[ \DYli[\alpha]{\gamma} G^{\gamma\beta}_{eHD4} + \Yli[\gamma]{\beta} G^{\ast\gamma\alpha}_{eHD4} \right]
\nonumber
\\
&=& \frac{g^2_1}{36} \delta^{\alpha\beta} \DDYni{i} \pM{-2} \left(5+6\lnmi\right) - \frac{1}{24} \left( \DYl \Yn \right)^{}_{\alpha i} \pM{-2} \left( \DYn \Yl \right)^{}_{i\beta} \left(5+6\lnmi\right) \;.
\label{eq:Hew}
\end{eqnarray}

\item $\psi^2H^3$
\begin{eqnarray}
C^{\alpha\beta}_{UH} &=& \Yui{\beta} \left( g^{}_2 G^{}_{WDH} + 2\lambda G^{}_{DH} + \frac{1}{2} G^\prime_{HD} -\rmI G^{\prime\prime}_{HD} \right)
\nonumber
\\
&=& \Yui{\beta} \left[ \frac{2\lambda}{3} {\rm tr} \left( \Yn M^{-2} \DYn \right) -\frac{g^2_2}{36} \DDYni{i} \pM{-2} \left(5+6\lnmi\right) + \frac{1}{4} \left( \DYn \Yl \DYl \Yn \right)^{}_{ii}  \right.
\nonumber
\\
&& \times \pM{-2} \left(1+2\lnmi\right) - \frac{1}{2} \DDYni{k} \DDYni[k]{i} \frac{\lnik}{\pM{2} - \pM[k]{2}} - \frac{1}{2} \DDYni{k} \DDYni{k}
\nonumber
\\
&& \times \left. \frac{\pM{4}(1+\lnmk) - \pM{2}\pM[k]{2} \left(1+2\lnmk\right) + \pM[k]{4} \lnmi }{\pM{}\pM[k]{}\left(\pM{2} - \pM[k]{2} \right)^2} \right] \;,
\label{eq:UHw}
\\
C^{\alpha\beta}_{dH} &=& \Ydi{\beta} \left( g^{}_2 G^{}_{WDH} + 2\lambda G^{}_{DH} + \frac{1}{2} G^\prime_{HD} +\rmI G^{\prime\prime}_{HD} \right)
\nonumber
\\
&=& \Ydi{\beta} \left[ \frac{2\lambda}{3} {\rm tr} \left( \Yn M^{-2} \DYn \right) -\frac{g^2_2}{36} \DDYni{i} \pM{-2} \left(5+6\lnmi\right) + \frac{1}{4} \left( \DYn \Yl \DYl \Yn \right)^{}_{ii}  \right.
\nonumber
\\
&& \times \pM{-2} \left(1+2\lnmi\right) - \frac{1}{2} \DDYni{k} \DDYni[k]{i} \frac{\lnik}{\pM{2} - \pM[k]{2}} - \frac{1}{2} \DDYni{k} \DDYni{k}
\nonumber
\\
&& \times \left. \frac{\pM{4}\lnmk - \pM{2}\pM[k]{2} \left(1+2\lnmi\right) + \pM[k]{4} \left(1+\lnmi\right) }{\pM{}\pM[k]{}\left(\pM{2} - \pM[k]{2} \right)^2} \right] \;,
\label{eq:DHw}
\end{eqnarray}
\begin{eqnarray}
C^{\alpha\beta}_{eH} &=& G^{\alpha\beta}_{eH} + \Yli{\beta} \left( g^{}_2  G^{}_{WDH} + 2\lambda G^{}_{DH} + \frac{1}{2} G^\prime_{HD} + \rmI G^{\prime\prime}_{HD} \right) - \frac{1}{2} \left(\Yl \DYl\right)^{}_{\alpha\gamma} \Yli[\lambda]{\beta} G^{\gamma\lambda}_{\ell D}
\nonumber
\\
&& - 2\lambda G^{\alpha\beta}_{eHD1} + \frac{1}{4} \left[ \left( \DYl \Yl \right)^{}_{\gamma\beta} G^{\alpha\gamma}_{eHD2} + \left( \Yl \DYl \right)^{}_{\alpha\gamma} G^{\gamma \beta}_{eHD2} - 2 \Yli{\gamma} \Yli[\lambda]{\beta} G^{\ast\lambda\gamma}_{eHD2} \right]
\nonumber
\\
&& - \lambda G^{\alpha\beta}_{eHD2} - \frac{1}{2} \left[ \left( \DYl \Yl \right)^{}_{\gamma\beta} G^{\alpha\gamma}_{eHD3} + \Yli{\gamma} \Yli[\lambda]{\beta} G^{\ast\lambda\gamma}_{eHD3} \right] + \frac{1}{4} \left[ \left( \DYl \Yl \right)^{}_{\gamma\beta} G^{\alpha\gamma}_{eHD4} \right.
\nonumber
\\
&& - \left. \left(\Yl \DYl\right)^{}_{\alpha\gamma} G^{\gamma\beta}_{eHD4} \right] + \lambda G^{\alpha\beta}_{eHD4} + \Yli[\gamma]{\beta} \left( G^{\prime(1)\alpha\gamma}_{H\ell} + G^{\prime(3)\alpha\gamma}_{H\ell}  + \rmI G^{\prime\prime(1)\alpha\gamma}_{H\ell}  \right) + \Yli{\gamma} G^{\prime \gamma\beta}_{He}
\nonumber
\\
&=& \Yli{\beta} \left[ \frac{2\lambda}{3} {\rm tr} \left( \Yn M^{-2} \DYn \right) - \frac{g^2_2}{36} \DDYni{i} \pM{-2} \left(5+6\lnmi\right) + \frac{1}{4} \left( \DYn \Yl \DYl \Yn \right)^{}_{ii} \right.
\nonumber
\\
&& \times \pM{-2} \left(1+2\lnmi\right) - \frac{1}{2} \DDYni{k} \DDYni[k]{i} \frac{\lnik}{\pM{2} - \pM[k]{2}} - \frac{1}{2} \DDYni{k} \DDYni{k}
\nonumber
\\
&& \times \left. \frac{\pM{4} \lnmk - \pM{2}\pM[k]{2} \left(1+2\lnmi\right) + \pM[k]{4} \left(1+\lnmi\right) }{\pM{}\pM[k]{} \left( \pM{2} - \pM[k]{2} \right)^2} \right] - \frac{1}{3} \left( \Yl\DYl\Yn M^{-2} \DYn\Yl \right)^{}_{\alpha\beta}
\nonumber
\\
&& - \frac{\lambda}{2} \Yni{i} \pM{-2} \left( \DYn\Yl \right)^{}_{i\beta} \left(3+2\lnmi\right) + \frac{1}{8} \Yni{i} \pM{-2} \left( \DYn\Yl\DYl\Yl\right)^{}_{i\beta} \left(3+2\lnmi\right)
\nonumber
\\
&&  + \frac{\Yni{i} \left( Y^{\rm T}_\nu Y^\ast_\nu \right)^{}_{ik} \left( \DYn \Yl \right)^{}_{k\beta}}{8\pM{}\pM[k]{} \left( \pM{2} - \pM[k]{2} \right)^2 } \left[ 3\pM{4}\left(1+2\lnmk\right) - 2\pM{2}\pM[k]{2} \left(5+\lnmk + 5\lnmi\right) \right.
\nonumber
\\
&& + \left. \pM[k]{4} \left(7+6\lnmi\right) \right] -\frac{1}{2} \Yni{i} \DDYni{k} \left(\DYn\Yl\right)^{}_{k\beta} \frac{\pM{2}\left(1-2\lnik\right) - \pM[k]{2} \left(1- \lnik\right)}{\left( \pM{2} - \pM[k]{2} \right)^2} \;.\qquad\quad
\label{eq:eHw}
\end{eqnarray}

\item Four-quark
\begin{eqnarray}
C^{(1)\alpha\beta\gamma\lambda}_{QU} &=& -\frac{1}{6} \Yui{\lambda} \DYui[\gamma]{\beta} G^{}_{DH} = -\frac{1}{18} {\rm tr} \left( \Yn M^{-2} \DYn \right) \Yui{\lambda} \DYui[\gamma]{\beta} \;,
\label{eq:QU1w}
\\
C^{(8)\alpha\beta\gamma\lambda}_{QU} &=& - \Yui{\lambda} \DYui[\gamma]{\beta} G^{}_{DH} = -\frac{1}{3} {\rm tr} \left( \Yn M^{-2} \DYn \right) \Yui{\lambda} \DYui[\gamma]{\beta} \;,
\label{eq:QU8w}
\\
C^{(1)\alpha\beta\gamma\lambda}_{Qd} &=& -\frac{1}{6} \Ydi{\lambda} \DYdi[\gamma]{\beta} G^{}_{DH} = -\frac{1}{18} {\rm tr} \left( \Yn M^{-2} \DYn \right) \Ydi{\lambda} \DYdi[\gamma]{\beta} \;,
\label{eq:Qd1w}
\\
C^{(8)\alpha\beta\gamma\lambda}_{Qd} &=& - \Ydi{\lambda} \DYdi[\gamma]{\beta} G^{}_{DH} = -\frac{1}{3} {\rm tr} \left( \Yn M^{-2} \DYn \right) \Ydi{\lambda} \DYdi[\gamma]{\beta} \;,
\label{eq:Qd8w}
\\
C^{(1)\alpha\beta\gamma\lambda}_{QUQd} &=& \Yui{\beta} \Ydi[\gamma]{\lambda} G^{}_{DH} = \frac{1}{3} {\rm tr} \left( \Yn M^{-2} \DYn \right) \Yui{\beta} \Ydi[\gamma]{\lambda} \;.
\label{eq:QUQdw}
\end{eqnarray}

\item Four-lepton
\begin{eqnarray}
C^{\alpha\beta\gamma\lambda}_{\ell\ell} &=& G^{\alpha\beta\gamma\lambda}_{\ell\ell} - \frac{g^{}_1}{2} \delta^{\gamma\lambda} G^{\alpha\beta}_{B\ell} + \frac{g^{}_2}{2} \left( 2\delta^{\gamma\beta} G^{\alpha\lambda}_{W\ell} - \delta^{\gamma\lambda} G^{\alpha\beta}_{W\ell} \right)
\nonumber
\\
&=& \frac{g^2_2 - g^2_1}{144} \delta^{\gamma\lambda} \Yni{i} \pM{-2} \DYni{i} \left(11+6\lnmi\right) - \frac{g^2_2}{72} \delta^{\gamma\beta} \Yni{i} \pM{-2} \DYni[\lambda]{i} \left(11+6\lnmi\right)
\nonumber
\\
&& -\frac{1}{8} \Yni{i} \pM{-2} \left[ \DYni{i} \left(\Yl\DYl\right)^{}_{\gamma\lambda} - \DYni[\lambda]{i} \left( \Yl\DYl \right)^{}_{\gamma\beta} \right] \left(3+2\lnmi\right)
\nonumber
\\
&& -\frac{1}{8} \Yni{i} \DYni[\lambda]{i} \Yni[\gamma]{k} \DYni[\beta]{k} \frac{\lnik}{ \pM{2} - \pM[k]{2} } -\frac{1}{4} \Yni{k} \Yni[\gamma]{k} \DYni[\beta]{i} \DYni[\lambda]{i}
\nonumber
\\
&& \times \frac{\pM{2} \left(1+\lnmk\right) - \pM[k]{2} \left(1+\lnmi\right) }{\pM{}\pM[k]{} \left( \pM{2} - \pM[k]{2} \right) } \;,
\label{eq:llw}
\\
C^{\alpha\beta\gamma\lambda}_{\ell e} &=& G^{\alpha\beta\gamma\lambda}_{\ell e} - \frac{1}{2} \Yli{\lambda} \DYli[\gamma]{\beta} G^{}_{DH} - g^{}_1 \delta^{\gamma\lambda} G^{\alpha\beta}_{B\ell} + \frac{1}{2} \left[ \DYli[\gamma]{\beta} G^{\alpha\lambda}_{eHD1} + \Yli{\lambda} G^{\ast\beta\gamma}_{eHD1} \right]
\nonumber
\\
&& + \frac{1}{4} \left[ \DYli[\gamma]{\beta} G^{\alpha\lambda}_{eHD2} + \Yli{\lambda} G^{\ast\beta\gamma}_{eHD2} \right] - \frac{1}{4} \left[ \DYli[\gamma]{\beta} G^{\alpha\lambda}_{eHD4} + \Yli{\lambda} G^{\ast\beta\gamma}_{eHD4} \right]
\nonumber
\\
&=&  -\frac{1}{6} {\rm tr} \left( \Yn M^{-2} \DYn \right) \Yli{\lambda} \DYli[\gamma]{\beta} + \frac{1}{4} \DYli[\gamma]{\beta} \left( \Yn M^{-2} \DYn \Yl \right)^{}_{\alpha\lambda}
\nonumber
\\
&&+ \frac{1}{4} \Yli{\lambda} \left( \DYl \Yn M^{-2} \DYn \right)^{}_{\gamma\beta} + \frac{1}{8} \Yni{i} \pM{-2} \DYni{i} \left(\DYl\Yl\right)^{}_{\gamma\lambda} \left(3+2\lnmi\right)
\nonumber
\\
&& -\frac{g^2_1}{72} \delta^{\gamma\lambda} \Yni{i} \pM{-2} \DYni{i} \left(11+6\lnmi\right) \;.\qquad\;
\label{eq:lew}
\end{eqnarray}

\item Semileptonic
\begin{eqnarray}
C^{(1)\alpha\beta\gamma\lambda}_{\ell Q} &=& G^{(1)\alpha\beta\gamma\lambda}_{\ell Q} + \frac{g^{}_1}{6} \delta^{\gamma\lambda} G^{\alpha\beta}_{B\ell}
\nonumber
\\
&=& \frac{1}{16} \Yni{i} \pM{-2} \DYni{i} \left[ \frac{g^2_1}{27} \delta^{\gamma\lambda} \left(11+6\lnmi\right) +  \left( \Yu\DYu - \Yd\DYd \right)^{}_{\gamma\lambda} \left(3+2\lnmi\right) \right] \;,
\label{eq:lQ1w}
\\
C^{(3)\alpha\beta\gamma\lambda}_{\ell Q} &=& G^{(3)\alpha\beta\gamma\lambda}_{\ell Q} + \frac{g^{}_2}{2} \delta^{\gamma\lambda} G^{\alpha\beta}_{W\ell}
\nonumber
\\
&=& \frac{1}{16} \Yni{i} \pM{-2} \DYni{i} \left[ -\frac{g^2_2}{9} \delta^{\gamma\lambda} \left(11+6\lnmi\right) + \left( \Yu\DYu + \Yd\DYd \right)^{}_{\gamma\lambda} \left(3+2\lnmi\right) \right] \;,\qquad
\label{eq:lQ3w}
\\
C^{\alpha\beta\gamma\lambda}_{\ell U} &=& G^{\alpha\beta\gamma\lambda}_{\ell U} + \frac{2g^{}_1}{3} \delta^{\gamma\lambda} G^{\alpha\beta}_{B\ell}
\nonumber
\\*
&=& \frac{1}{8} \Yni{i} \pM{-2} \DYni{i} \left[ \frac{2g^2_1}{27} \delta^{\gamma\lambda} \left(11+6\lnmi\right) -  \left( \DYu\Yu \right)^{}_{\gamma\lambda} \left(3+2\lnmi\right) \right] \;,
\label{eq:lUw}
\\
C^{\alpha\beta\gamma\lambda}_{\ell d} &=& G^{\alpha\beta\gamma\lambda}_{\ell d} - \frac{g^{}_1}{3} \delta^{\gamma\lambda} G^{\alpha\beta}_{B\ell}
\nonumber
\\
&=& \frac{1}{8} \Yni{i} \pM{-2} \DYni{i} \left[ -\frac{g^2_1}{27} \delta^{\gamma\lambda} \left(11+6\lnmi\right) + \left( \DYd\Yd \right)^{}_{\gamma\lambda} \left(3+2\lnmi\right) \right] \;,
\label{eq:ldw}
\\
C^{\alpha\beta\gamma\lambda}_{\ell e d Q} &=& \DYdi[\gamma]{\lambda} \left[ \Yli{\beta}  G^{}_{DH} - G^{\alpha\beta}_{eHD1} - \frac{1}{2} G^{\alpha\beta}_{eHD2} + \frac{1}{2} G^{\alpha\beta}_{eHD4}  \right]
\nonumber
\\
&=& \DYdi[\gamma]{\lambda} \left[ \frac{1}{3} {\rm tr} \left(\Yn M^{-2} \DYn \right) \Yli{\beta} - \frac{1}{2} \left( \Yn M^{-2} \DYn \Yl \right)^{}_{\alpha\beta} \right] \;,
\label{eq:ledQw}
\\
C^{(1)\alpha\beta\gamma\lambda}_{\ell e Q U} &=& \Yui[\gamma]{\lambda} \left[ -\Yli{\beta}  G^{}_{DH} + G^{\alpha\beta}_{eHD1} + \frac{1}{2} G^{\alpha\beta}_{eHD2} - \frac{1}{2} G^{\alpha\beta}_{eHD4}  \right]
\nonumber
\\
&=& \Yui[\gamma]{\lambda} \left[ -\frac{1}{3} {\rm tr} \left(\Yn M^{-2} \DYn \right) \Yli{\beta} + \frac{1}{2} \left( \Yn M^{-2} \DYn \Yl \right)^{}_{\alpha\beta} \right] \;.
\label{eq:leQUw}
\end{eqnarray}

\end{itemize}

\subsection{Simplified results}

To simplify the above results and make them more illuminating, we now assume that the masses of heavy Majorana neutrinos are exactly degenerate, i.e., $M^{}_1=M^{}_2=M^{}_3=M$ and take the matching scale to be $\mu=M$. All the divergences can be simply dropped in the $\overline{\rm MS}$ scheme.
Thus for the effective couplings and the Wilson coefficient of the dimension-five operator, we have
\begin{eqnarray}
m^2_{\rm eff} &=& m^2 + \frac{1}{\left(4\pi\right)^2} \left( 2 M^2-\frac{m^2}{2} - \frac{m^4}{3M^2} \right) {\rm tr} \left( \DYn \Yn \right) \;,
\\
\lambda_{\rm eff} &=& \lambda + \frac{1}{\left(4\pi\right)^2} \left[  \frac{5 g^2_2 m^2 - 18\lambda M^2 - 24 \lambda m^2}{18M^2} {\rm tr} \left( \DYn \Yn \right) + \frac{m^2}{M^2} {\rm tr} \left( \DYn \Yn \DYn \Yn \right) \right.
\nonumber
\\
&& - \left. \frac{2M^2-3m^2}{2M^2} {\rm tr} \left( \DYn \Yn Y^{\rm T}_\nu Y^\ast_\nu \right) - \frac{m^2}{2M^2} {\rm tr} \left( \DYn \Yl \DYl \Yn \right) \right] \;,
\\
Y^{\rm eff}_l &=& Y^{}_l - \frac{1}{\left(4\pi\right)^2} \left[ \left( \frac{1}{4} + \frac{m^2}{3M^2} \right) {\rm tr} \left( \DYn \Yn \right) Y^{}_l - \left( \frac{5}{8} + \frac{3m^2}{4M^2} \right) \Yn \DYn \Yl \right] \;,
\\
Y^{\rm eff}_{\rm u} &=& Y^{}_{\rm u} - \frac{1}{\left( 4\pi\right)^2} \left( \frac{1}{4} + \frac{m^2}{3M^2} \right) {\rm tr} \left( \DYn \Yn \right) Y^{}_{\rm u} \;,
\\
Y^{\rm eff}_{\rm d} &=& Y^{}_{\rm u} - \frac{1}{\left( 4\pi\right)^2} \left( \frac{1}{4} + \frac{m^2}{3M^2} \right) {\rm tr} \left( \DYn \Yn \right) Y^{}_{\rm d} \;,
\\
C^{(5)}_{\rm eff} &=& C^{(5)} + \frac{1}{\left( 4\pi\right)^2} \left\{\hspace{-0.1cm} \left[ 2\lambda + \frac{ g^2_1 + g^2_2}{4}  - \frac{{\rm tr} \left( \DYn \Yn \right)}{2}  \right] C^{(5)} -  \frac{3}{8} \left( C^{(5)} Y^\ast_\nu Y^{\rm T}_\nu + \Yn \DYn C^{(5)} \right) \hspace{-0.1cm}\right\} \;,\quad
\end{eqnarray}
where $ C^{(5)} = \Yn Y^{\rm T}_\nu/M$ is the tree-level contribution to the Wilson coefficient of the unique dimension-five operator. The tree-level contributions to the Wilson coefficients of $\Op^{(1)\alpha\beta}_{H\ell}$ and $\Op^{(3)\alpha\beta}_{H\ell}$ are given by
\begin{eqnarray}
\left[ \Op^{(1)}_{H\ell} \right]^{\alpha\beta}_{\rm tree} = - \left[ \Op^{(3)}_{H\ell} \right]^{\alpha\beta}_{\rm tree} = \frac{1}{4M^2} \left( \Yn\DYn \right)^{}_{\alpha\beta} \;.
\end{eqnarray}
Again, the one-loop-level Wilson coefficients of the operators listed in Table~\ref{tb:Wbasis} should be multiplied by an overall loop factor $1/\left( 4\pi \right)^2$, which will not be explicitly shown in all the Wilson coefficients. Under the assumption of mass degeneracy, the simplified Wilson coefficients can be found below:

\begin{itemize}
\item $X^2H^2$
\begin{eqnarray}
C^{}_{HB} &=& \frac{g^2_1}{24M^2} {\rm tr} \left( \DYn \Yn \right) \;,
\\
C^{}_{HWB} &=& \frac{g^{}_1g^{}_2}{12M^2} {\rm tr} \left( \DYn \Yn \right) \;,
\\
C^{}_{HW} &=& \frac{g^2_2}{24M^2} {\rm tr} \left( \DYn \Yn \right) \;.
\end{eqnarray}

\item $H^4D^2$
\begin{eqnarray}
C^{}_{H\square} &=& - \frac{1}{72M^2} \left[ 5\left( g^2_1 + 3g^2_2 \right) {\rm tr} \left( \DYn \Yn \right) + 24 {\rm tr} \left( \DYn \Yn \DYn \Yn \right) + 78 {\rm tr} \left( \DYn \Yn Y^{\rm T}_\nu Y^\ast_\nu \right) \right.
\nonumber
\\
&& - \left. 18 {\rm tr} \left( \DYn \Yl \DYl \Yn  \right) \right] \;,
\\
C^{}_{HD} &=& - \frac{1}{18M^2} \left[ 5g^2_1 {\rm tr} \left( \DYn \Yn \right) + 9 {\rm tr} \left( \DYn \Yn \DYn \Yn \right) + 36 {\rm tr} \left( \DYn \Yn Y^{\rm T}_\nu Y^\ast_\nu \right) \right.
\nonumber
\\
&& + \left. 9 {\rm tr} \left( \DYn \Yl \DYl \Yn  \right) \right] \;.
\end{eqnarray}

\item $H^6$
\begin{eqnarray}
C^{}_H &=& \frac{1}{M^2} \left\{ \lambda  \left[ \frac{\left( 12\lambda - 5g^2_2 \right)}{9} {\rm tr} \left( \DYn \Yn \right) - 2 {\rm tr} \left( \DYn \Yn \DYn \Yn \right) - 3 {\rm tr} \left( \DYn \Yn Y^{\rm T}_\nu Y^\ast_\nu \right) \right.\right.
\nonumber
\\
&& + \left.\left. {\rm tr} \left( \DYn \Yl \DYl \Yn  \right) \right] + \frac{1}{3} {\rm tr} \left( \DYn\Yn\DYn\Yn\DYn\Yn \right) - {\rm tr} \left( \DYn\Yn\DYn\Yn Y^{\rm T}_\nu Y^\ast_\nu \right) \right\} \;.
\end{eqnarray}

\item $\psi^2 X H$
\begin{eqnarray}
C^{\alpha\beta}_{eB} &=& \frac{g^{}_1}{24M^2} \left( \Yn \DYn \Yl \right)^{}_{\alpha\beta} \;,
\\
C^{\alpha\beta}_{eW} &=& \frac{5g^{}_2}{24M^2} \left( \Yn \DYn \Yl \right)^{}_{\alpha\beta} \;.
\end{eqnarray}

\item $\psi^2DH^2$
\begin{eqnarray}
C^{(1)\alpha\beta}_{HQ} &=& - \frac{5 \delta^{\alpha\beta} g^2_1}{216M^2} {\rm tr} \left( \DYn\Yn \right) \;,
\\
C^{(3)\alpha\beta}_{HQ} &=& - \frac{5 \delta^{\alpha\beta} g^2_2}{72M^2} {\rm tr} \left( \DYn\Yn \right) \;,
\\
C^{\alpha\beta}_{HU} &=& - \frac{5 \delta^{\alpha\beta} g^2_1}{54M^2} {\rm tr} \left( \DYn\Yn \right) \;,
\\
C^{\alpha\beta}_{Hd} &=&  \frac{5 \delta^{\alpha\beta} g^2_1}{108M^2} {\rm tr} \left( \DYn\Yn \right) \;,
\\
C^{(1)\alpha\beta}_{H\ell} &=& \frac{1}{M^2} \left\{ \frac{5\delta^{\alpha\beta} g^2_1}{72} {\rm tr} \left( \DYn \Yn \right) + \left[ \frac{11}{288} \left( 11g^2_1 + 27 g^2_2 \right) - \frac{1}{8} {\rm tr} \left( \DYn \Yn \right) \right] \left( \Yn \DYn \right)^{}_{\alpha\beta} \right.
\nonumber
\\
&& - \frac{11}{16} \left( \Yn\DYn\Yn\DYn \right)^{}_{\alpha\beta} - \frac{1}{2} \left( \Yn Y^{\rm T}_\nu Y^\ast_\nu \DYn \right)^{}_{\alpha\beta} - \frac{3}{16} \left( \Yl\DYl\Yn\DYn \right)^{}_{\alpha\beta}
\nonumber
\\
&& - \left. \frac{3}{16} \left( \Yn\DYn\Yl\DYl \right)^{}_{\alpha\beta} \right\} \;,
\\
C^{(3)\alpha\beta}_{H\ell}  &=& \frac{1}{M^2} \left\{ - \frac{5\delta^{\alpha\beta} g^2_2}{72} {\rm tr} \left( \DYn \Yn \right) + \left[ \frac{11}{288} \left( -9g^2_1 + 7 g^2_2 \right) + \frac{1}{8} {\rm tr} \left( \DYn \Yn \right) \right] \left( \Yn \DYn \right)^{}_{\alpha\beta} \right.
\nonumber
\\
&& + \frac{3}{16} \left( \Yn\DYn\Yn\DYn \right)^{}_{\alpha\beta} + \frac{1}{8} \left( \Yn Y^{\rm T}_\nu Y^\ast_\nu \DYn \right)^{}_{\alpha\beta} - \frac{3}{16} \left( \Yl\DYl\Yn\DYn \right)^{}_{\alpha\beta}
\nonumber
\\
&& - \left. \frac{3}{16} \left( \Yn\DYn\Yl\DYl \right)^{}_{\alpha\beta} \right\} \;,
\\
C^{\alpha\beta}_{He} &=&  \frac{5}{72M^2} \left[ 2 \delta^{\alpha\beta} g^2_1 {\rm tr} \left( \DYn \Yn \right) - 3 \left( \DYl \Yn \DYn \Yl \right)^{}_{\alpha\beta} \right] \;.
\end{eqnarray}

\item $\psi^2 H^3$
\begin{eqnarray}
C^{\alpha\beta}_{UH} &=& \frac{1}{36M^2} \left( Y^{}_{\rm u} \right)^{}_{\alpha\beta} \left[ \left( 24\lambda - 5 g^2_2 \right) {\rm tr} \left( \DYn \Yn \right) - 18 {\rm tr} \left( \DYn\Yn\DYn\Yn \right) - 27 {\rm tr} \left( \DYn\Yn Y^{\rm T}_\nu Y^\ast_\nu \right) \right.
\nonumber
\\
&& + \left.  9{\rm tr} \left( \Yn\DYn\Yl\DYl \right) \right] \;,
\\
C^{\alpha\beta}_{dH} &=& \frac{1}{36M^2} \left( Y^{}_{\rm d} \right)^{}_{\alpha\beta} \left[ \left( 24\lambda - 5 g^2_2 \right) {\rm tr} \left( \DYn \Yn \right) - 18 {\rm tr} \left( \DYn\Yn\DYn\Yn \right) - 27 {\rm tr} \left( \DYn\Yn Y^{\rm T}_\nu Y^\ast_\nu \right) \right.
\nonumber
\\
&& + \left. 9{\rm tr} \left( \Yn\DYn\Yl\DYl \right) \right] \;,
\\
C^{\alpha\beta}_{eH} &=& \frac{1}{36M^2} \left( Y^{}_l \right)^{}_{\alpha\beta} \left[ \left( 24\lambda - 5 g^2_2 \right) {\rm tr} \left( \DYn \Yn \right) - 18 {\rm tr} \left( \DYn\Yn\DYn\Yn \right) - 27 {\rm tr} \left( \DYn\Yn Y^{\rm T}_\nu Y^\ast_\nu \right) \right.
\nonumber
\\
&& + \left. 9{\rm tr} \left( \Yn\DYn\Yl\DYl \right) \right] + \frac{1}{M^2} \left[ -\frac{3\lambda}{2} \left( \Yn\DYn\Yl \right)^{}_{\alpha\beta} + \frac{3}{4} \left( \Yn\DYn\Yn\DYn\Yl \right)^{}_{\alpha\beta} \right.
\nonumber
\\
&& + \left. \frac{11}{8} \left( \Yn Y^{\rm T}_\nu Y^\ast_\nu \DYn \Yl \right)^{}_{\alpha\beta} - \frac{1}{3} \left( \Yl\DYl\Yn\DYn\Yl \right)^{}_{\alpha\beta} + \frac{3}{8} \left( \Yn\DYn\Yl\DYl\Yl \right)^{}_{\alpha\beta} \right] \;.
\end{eqnarray}

\item Four-quark
\begin{eqnarray}
C^{(1)\alpha\beta\gamma\lambda}_{QU} &=& - \frac{1}{18M^2} {\rm tr} \left(\DYn\Yn \right) \left( \Yu \right)^{}_{\alpha\lambda} \left( \DYu \right)^{}_{\gamma\beta} \;,
\\
C^{(8)\alpha\beta\gamma\lambda}_{QU} &=& - \frac{1}{3M^2} {\rm tr} \left(\DYn\Yn \right) \left( \Yu \right)^{}_{\alpha\lambda} \left( \DYu \right)^{}_{\gamma\beta} \;,
\\
C^{(1)\alpha\beta\gamma\lambda}_{Qd} &=& - \frac{1}{18M^2} {\rm tr} \left(\DYn\Yn \right) \left( \Yd \right)^{}_{\alpha\lambda} \left( \DYd \right)^{}_{\gamma\beta} \;,
\\
C^{(8)\alpha\beta\gamma\lambda}_{Qd} &=& - \frac{1}{3M^2} {\rm tr} \left(\DYn\Yn \right) \left( \Yd \right)^{}_{\alpha\lambda} \left( \DYd \right)^{}_{\gamma\beta} \;,
\\
C^{(1)\alpha\beta\gamma\lambda}_{QUQd} &=& \frac{1}{3M^2} {\rm tr} \left(\DYn\Yn \right) \left( \Yu \right)^{}_{\alpha\beta} \left( \Yd \right)^{}_{\gamma\lambda} \;.
\end{eqnarray}

\item Four-lepton
\begin{eqnarray}
C^{\alpha\beta\gamma\lambda}_{\ell\ell} &=& \frac{1}{M^2} \left[ \frac{11\left(g^2_2-g^2_1 \right)}{144} \delta^{\gamma\lambda} \left( \Yn\DYn \right)^{}_{\alpha\beta} - \frac{11g^2_2}{72} \delta^{\gamma\beta} \left( \Yn\DYn \right)^{}_{\alpha \lambda} - \frac{1}{8} \left( \Yn\DYn\right)^{}_{\alpha\lambda} \left( \Yn\DYn\right)^{}_{\gamma\beta} \right.
\nonumber
\\
&& - \left. \frac{1}{2} \left( \Yn Y^{\rm T}_\nu \right)^{}_{\alpha\gamma} \left( Y^\ast_\nu \DYn \right)^{}_{\beta\lambda} - \frac{3}{8} \left( \Yn\DYn \right)^{}_{\alpha\beta} \left(\Yl\DYl\right)^{}_{\gamma\lambda} + \frac{3}{8} \left( \Yn\DYn \right)^{}_{\alpha\lambda} \left( \Yl\DYl \right)^{}_{\gamma\beta} \right] \;,\quad
\\
C^{\alpha\beta\gamma\lambda}_{\ell e} &=& - \frac{1}{M^2} \left[ \frac{11g^2_1}{72} \delta^{\gamma\lambda} \left( \Yn\DYn \right)^{}_{\alpha\beta} + \frac{1}{6} {\rm tr} \left(\DYn\Yn\right) \Yli{\lambda}\DYli[\gamma]{\beta} - \frac{1}{4} \left( \Yn\DYn\Yl \right)^{}_{\alpha\lambda} \DYli[\gamma]{\beta} \right. \quad
\nonumber
\\
&&  - \left. \frac{3}{8} \left(\Yn\DYn\right)^{}_{\alpha\beta} \left( \DYl\Yl \right)^{}_{\gamma\lambda} - \frac{1}{4} \Yli{\lambda} \left(\DYl\Yn\DYn\right)^{}_{\gamma\beta} \right] \;.
\end{eqnarray}

\item Semileptonic
\begin{eqnarray}
C^{(1)\alpha\beta\gamma\lambda}_{\ell Q} &=& \frac{1}{16M^2} \left( \Yn\DYn \right)^{}_{\alpha\beta} \left[ \frac{11g^2_1}{27} \delta^{\gamma\lambda} + 3 \left( \Yu\DYu - \Yd\DYd \right)^{}_{\gamma\lambda} \right] \;,
\\
C^{(3)\alpha\beta\gamma\lambda}_{\ell Q} &=& \frac{1}{16M^2} \left( \Yn\DYn \right)^{}_{\alpha\beta} \left[ - \frac{11g^2_2}{9} \delta^{\gamma\lambda} + 3 \left( \Yu\DYu + \Yd\DYd \right)^{}_{\gamma\lambda} \right] \;,
\\
C^{\alpha\beta\gamma\lambda}_{\ell U} &=& \frac{1}{8M^2} \left( \Yn\DYn \right)^{}_{\alpha\beta} \left[ \frac{22g^2_1}{27} \delta^{\gamma\lambda} - 3 \left( \DYu\Yu \right)^{}_{\gamma\lambda} \right] \;,
\\*
C^{\alpha\beta\gamma\lambda}_{\ell d} &=& \frac{1}{8M^2} \left( \Yn\DYn \right)^{}_{\alpha\beta} \left[ - \frac{11g^2_1}{27} \delta^{\gamma\lambda} + 3 \left( \DYd\Yd \right)^{}_{\gamma\lambda} \right] \;,
\\
C^{\alpha\beta\gamma\lambda}_{\ell edQ} &=& \frac{1}{M^2} \left( \DYd \right)^{}_{\gamma\lambda} \left[ \frac{1}{3} {\rm tr} \left( \DYn\Yn \right) \Yli{\beta} - \frac{1}{2} \left( \Yn\DYn\Yl \right)^{}_{\alpha\beta} \right] \;,
\\
C^{\alpha\beta\gamma\lambda}_{\ell eQU} &=& \frac{1}{M^2} \left( \Yu \right)^{}_{\gamma\lambda} \left[ - \frac{1}{3} {\rm tr} \left( \DYn\Yn \right) \Yli{\beta} + \frac{1}{2} \left( \Yn\DYn\Yl \right)^{}_{\alpha\beta} \right] \;.
\end{eqnarray}
\end{itemize}
Further simplifications could be made by noticing the strong hierarchy among the SM fermion Yukawa couplings. Since this can be easily achieved by just ignoring the relatively small Yukawa couplings, it is unnecessary to do this explicitly.

Note that the redundant operators in the Green's basis have been removed via the EOMs of relevant fields, whereas new independent operators appear in the Warsaw basis. Though the heavy Majorana neutrinos do not directly interact with quarks in the type-I seesaw model, the pure quark interactions (i.e., the four-quark operators) are present. With Eqs.~(\ref{eq:QU1w})-(\ref{eq:QUQdw}), one can easily find that all these four-quark operators result from the $\Op^{}_{DH}$ in the Green's basis. Therefore, the complete Lagrangian of the SEFT up to the one-loop level reads
\begin{eqnarray}
\mathcal{L}^{}_{\rm SEFT} &=& \mathcal{L}^{}_{\rm SM} \left( m^2 \to m^2_{\rm eff}, \lambda \to \lambda^{}_{\rm eff}, Y^{}_l \to Y^{\rm eff}_l, Y^{}_{\rm u} \to Y^{\rm eff}_{\rm u}, Y^{}_{\rm d} \to Y^{\rm eff}_{\rm d}  \right)
\nonumber
\\
&& + \left[ \frac{1}{2} \left( C^{(5)}_{\rm eff} \right)_{\alpha\beta} \Op^{(5)}_{\alpha\beta} + {\rm h.c.} \right] + \frac{1}{4} C^{(6)}_{\alpha\beta} \left[ \Op^{(1)\alpha\beta}_{H\ell} - \Op^{(3)\alpha\beta}_{H\ell} \right] + \sum_i C^{}_i \Op^{}_{i} \;,
\end{eqnarray}
where $\Op^{}_i$ denote the dimension-six operators listed in Table~\ref{tb:Wbasis} including Hermitian conjugations of the non-Hermitian operators, while $C^{}_i$ refer to the one-loop contributions to corresponding Wilson coefficients. The coefficients $C^{}_i$ are suppressed both by the mass scale of the heavy Majorana neutrinos and by the loop factor, i.e., $1/M^2$ and $1/\left(16 \pi^2\right)$. Unlike what we have done for the dimension-five operator, the tree- and one-loop-level contributions to the Wilson coefficients of $\Op^{(1)\alpha\beta}_{H\ell}$ and $\Op^{(3)\alpha\beta}_{H\ell}$ are not summed up.

Before ending this section, we briefly discuss possible applications of the one-loop matching results. First, a self-consistent calculation of radiative decays of charged leptons in the SEFT has been performed in Ref.~\cite{Zhang:2021tsq} where the one-loop matching for the relevant operators is carried out via diagrammatic calculations. One can observe that the Wilson coefficients of the operators $\mathcal{O}^{\alpha\beta}_{\ell D}$, $\mathcal{O}^{\alpha\beta}_{eB}$, $\mathcal{O}^{\alpha\beta}_{eW}$, $\mathcal{O}^{\alpha\beta}_{eHD1}$, $\mathcal{O}^{\alpha\beta}_{eHD2}$, $\mathcal{O}^{\alpha\beta}_{eHD3}$, and $\mathcal{O}^{\alpha\beta}_{eHD4}$ given in Eqs.~(\ref{eq:lD})-(\ref{eq:eHD4}) in the Green's basis exactly coincide with those given in Eqs.~(11), (15) and (18) of Ref.~\cite{Zhang:2021tsq}, after $\mu = \mathcal{O} \left( M^{}_i \right)$ is taken and the different conventions for the last four operators are considered. Actually, one can check that if $\mu = \mathcal{O} \left( M^{}_i \right)$ is not taken during the matching done in Ref.~\cite{Zhang:2021tsq}, the results are exactly the same as those given in Eqs.~(\ref{eq:lD})-(\ref{eq:eHD4}) as they should be. After applying the EOMs of relevant fields, we obtain the Wilson coefficients of operators $\mathcal{O}^{\alpha\beta}_{eB}$ and $\mathcal{O}^{\alpha\beta}_{eW}$ in the Warsaw basis, as shown in Eqs.~(\ref{eq:eBw}) and (\ref{eq:eWw}). These results are identical to those in Eq. (23) of Ref.~\cite{Zhang:2021tsq}. Then, one may also make use of the above results to study the contributions from the heavy Majorana neutrinos to other leptonic observables. For instance, $\Op^{\alpha\beta}_{eH}$ contributes to $H \to \ell^-_\alpha\ell^+_\beta$ ($\alpha=\beta$ for decays into the same lepton flavor and $\alpha \neq \beta$ for lepton-flavor-violating decays), while $\Op^{(1) \alpha\beta}_{H\ell}$ and $\Op^{(3) \alpha\beta}_{H\ell}$ lead to $Z \to \nu^{}_\alpha \overline{\nu}^{}_\beta$ at both the tree and one-loop levels. But the latter make contributions to $Z \to \ell^-_\alpha \ell^+_\beta$ only at the one-loop level, together with $\Op^{\alpha\beta}_{He}$, $\Op^{\alpha\beta}_{eB}$ and $\Op^{\alpha\beta}_{eW}$, since at the tree level the combination $\Op^{(1) \alpha\beta}_{H\ell}- \Op^{(3) \alpha\beta}_{H\ell}$ just modifies the coupling of ordinary neutrinos to $W$ and $Z$ gauge bosons. Besides $\ell^-_\beta \to \ell^-_\alpha \gamma$, $\Op^{\alpha\beta}_{eB}$ and $\Op^{\alpha\beta}_{eW}$ also contribute to magnetic and electric dipole moments of charged leptons, while the semi-leptonic operators cause the $\mu$-$e$ conversion. The operators $\Op^{\alpha\beta\gamma\lambda}_{\ell\ell}$ and $\Op^{\alpha\beta\gamma\lambda}_{\ell e}$ imply $\mu \to eee$ and also $\ell \to \ell \overline{\nu} \nu$, resulting in a shift of the Fermi constant $G^{}_{\rm F}$. Finally, the pure- and semi-leptonic operators provide us with non-standard interactions of ordinary neutrinos, which could be probed in future neutrino oscillation experiments~\cite{Du:2020dwr, Du:2021rdg}.

\section{Summary}\label{sec:summary}
The compelling experimental evidence for neutrino masses and lepton flavor mixing indicates that the SM is incomplete and serves only as an EFT at the electroweak scale. The type-I seesaw model with three right-handed neutrinos is a simple and natural extension of the SM to accommodate tiny masses of active neutrinos. Since these right-handed neutrinos may be too heavy to be directly produced and detected in the terrestrial collider experiments, one usually explores their low-energy phenomenological consequences by using the precision measurements. For this purpose, it is useful to establish the low-energy EFT of the type-I seesaw model by integrating out the heavy Majorana neutrinos and thus their impact is encoded in the Wilson coefficients of higher-dimensional operators.

In this work, we have carried out the complete one-loop matching of the type-I seesaw model onto the SMEFT up to dimension-six operators by using the functional approach. First of all, such an investigation adds another example of a complete one-loop matching for a given UV model. We explain the generic framework of the functional approach to both the tree-level and the one-loop matching. After integrating out the heavy Majorana neutrinos, we obtain one dimension-five operator and one dimension-six operator at the tree level, which are well-known in the literature, and 31 independent dimension-six operators (barring flavor structure and Hermitian conjugates) in the Warsaw basis with the one-loop matching. This number is just about one half of that for the independent dimension-six operators in the SMEFT. The tree-level and one-loop contributions to the Wilson coefficients of these dimension-six operators are presented up to $\mathcal{O}\left( M^{-2} \right)$. It is worth pointing out that parts of the one-loop contributions to the Wilson coefficients of $\mathcal{O}^{(1)\alpha\beta}_{H\ell}$ and $\mathcal{O}^{(3)\alpha\beta}_{H\ell}$ come from the tree-level contributions via the redefinitions of $H$ and $\ell$. In addition, the one-loop threshold corrections to the couplings in the SM and also to the coefficient of the dimension-five operator are given up to $\mathcal{O}\left( M^{-2} \right)$, which are the very matching conditions for the two-loop RGEs of these couplings.

However, we have assumed that the masses of three heavy Majorana neutrinos are nearly degenerate and they can be integrated out simultaneously. If their masses are hierarchical, one should integrate out heavy Majorana neutrinos sequentially, and construct the EFT between any two mass scales and implement the RGEs to connect the physical parameters at those two mass scales. More explicitly, the Wilson coefficients of the operators are needed to be evolved down to the energy scale of relevant experiments from the matching scale via their RGEs. For a self-consistent calculation, the two-loop RGEs have to be derived, which is obviously out of the scope of this work. We leave the construction of the EFTs at the intermediate scale and two-loop RGEs in the SEFT for future works.

\section*{Acknowledgements}
This work was supported in part by the National Natural Science Foundation of China under grant No.~11775231, No.~11775232, No.~11835013 and No.~12075254, by the Key Research Program of the Chinese Academy of Sciences under grant No. XDPB15, and by the CAS Center for Excellence in Particle Physics.

\end{document}